\documentclass{aastex631}

\usepackage{amsmath}
\usepackage{mathtools}
\shorttitle{Cellular Dynamics of Helium-rich Detonation}
\shortauthors{Iwata and Maeda}
\graphicspath{{./}{figures/}}

\begin{document}

\title{Cellular Dynamics of Helium-rich Detonation on sub-Chandrasekhar Mass White Dwarf \footnote{Released on March, 1st, 2021}}

\author[0000-0001-6195-0032]{Kazuya Iwata}
\affiliation{Department of Mechanical Engineering and Science, Kyoto University \\
Kyotodaigaku-Katsura, Nishikyo-ku \\
Kyoto, Japan}

\author[0000-0003-2611-7269]{Keiichi Maeda}
\affiliation{Department of Astronomy, Kyoto University \\
Kitashirakawa-Oiwake-cho, Sakyo-ku, Kyoto 606-8502, Japan}




\begin{abstract}

Most previous efforts for hydrodynamic studies on detonation in the context of Type Ia supernovae did not take into account the scale of the cellular structure for a criterion in initiation, propagation, quenching, and the resolution requirement of detonation, whereas it is quite common to consider cell sizes in the discussion on terrestrial detonation in chemically reactive systems. In our recent study, the terrestrial cell-based theories, which incorporates the cell-size data acquired in 2D simulations of helium detonation in the double-detonation model, were demonstrated to be a powerful diagnostics in reproducing the thresholds in the initiation and quenching provided by previous studies. In the present study, 2D simulation results of the cellular detonation in the base of white-dwarf (WD) envelope are described in detail, in terms of the dynamic wave morphology and chemical abundance structure. The cellular structure is observed at a range of upstream density and envelope composition explored in the present work. C/O contamination by the WD core material reduces the cell width rapidly, as accelerated by the $\alpha$-capture reaction. It is also indicated that nickel production could be significantly delayed for the C/O-rich composition. The small cell width makes it extremely demanding to resolve the detonation structure in full-star simulations of SNe Ia; this could raise a concern on the robustness of the outcomes of some numerical simulations in terms of the success and failure of detonation. This issue may be overcome by sub-grid modeling that incorporates the cellular dynamics acquired in resolved simulations.

\end{abstract}

\keywords{Type Ia supernovae (1728) --- Hydrodynamical simulations (1736)}


\section{Introduction} \label{sec:intro}

Despite substantial effort dedicated to understanding the underlying mechanism of normal Type Ia supernovae (SNe Ia), consensus has not been reached so far on its progenitor system \citep{Branch_1995}. It is widely accepted that it is driven by the thermonuclear runaway of a mass-accreting white dwarf (WD) in a binary system. However, the textbook explosion model of a near-Chandrasekhar mass WD, which has long been believed to be the progenitor of normal SNe Ia, lacks key evidence of the non-degenerate companion star in the observations \citep[see, e.g.,][for a review]{Maeda2016}. In addition, discoveries of diverse classes of peculiar SNe Ia now strongly indicate multiple progenitor systems for SNe Ia \citep[e. g.,][]{Taubenberger2017}. 

Most of the explosion models of SNe Ia involve detonation, which is a supersonic regime of the thermonuclear combustion front. Hence, precise modeling of the detonation physics is key to estimating the plausibility of the explosion models. The detonation was originally discovered in terrestrial chemically reactive systems around the end of the 19th century, which has a close analogy to astrophysical thermonuclear detonation \citep{NOURI2019156,ORAN20051823}. Detonation in the terrestrial systems has been extensively studied up to date, originally for preventing hazardous explosion accidents in coal mines and chemical plants, and more recently for developing high-efficiency aerospace propulsion systems \citep{lee_2008,WOLANSKI2013125}. 

Deflagration, which is a subsonic counterpart of combustion phenomena driven by heat conduction and mass diffusion, is also described similarly in terrestrial and astrophysical systems. \cite{PoludnenkoScience2019} presented a key study in which deflagration-to-detonation transition (DDT) in the delayed-detonation model was discussed via terrestrial experiments and chemical detonation simulations. A faint class of peculiar SNe Ia known as SNe Iax has been suggested to be driven primarily by deflagration \citep{Foley_2013,Feldman2023}. Studies on these two distinct combustion regimes, deflagration and detonation, in the context of SNe Ia can be found in the abundance of literature \citep[e. g.,][]{PhysRevLett.92.211102,Khokhlov_1997,Townsley_2019,TimmesNiemeyer_2000,Brookeretal2021}.

One major problem in hydrodynamic simulations of thermonuclear flame is the numerical resolution; it is challenging to resolve detailed burning processes in full-star simulations. Resolutions of the order of km are common in the state-of-the-art multidimensional studies \citep[e.g.,][]{Rivas_2022,Roy_2022,Pakmor_2022}, whereas the spatial scales for fuel consumption and energy generation can reach the order of cm in high-density media. \cite{Gamezo_1999,Timmes_2000,Boisseau_1996,Papatheodore_2014} performed highly-resolved 2D simulations of the carbon/oxygen detonation, in which they needed the resolutions of $10^{-2}-10^3$ cm to capture the cellular structure, which is an essential feature of the detonation regardless of being chemical or thermonuclear.

The burning limiter proposed by \cite{Kushnir_2013} and \cite{Kushnir2020} is one of the useful techniques employed in recent studies to resolve this issue, in which the progress of nuclear reaction is limited so that detonation structure is broadened to resolved scales. It is close to the artificially thickened flame modeling in its concept \citep{Butler1977ANM}, which is often applied in terrestrial combustion problems to control the flame thickness without influencing its propagation speed. However, the burning limiter was reported to decrease the detonation velocity and affect the resulting chemical abundances \citep{Miles_2019}.

Some other studies addressed the resolution dependence of the detonation. \cite{Shen_2014co} performed 1D study to show how the success/failure of the detonation in the center of WD depends on the resolution. \cite{Rivas_2022} showed that the general explosion outcomes of the double-detonation model could be consistent with the resolutions of 0.5-2 km. \cite{Iwata_2022} performed 1D study on the self-ignition of helium-rich detonation in the WD envelope base by including mass accretion, and showed that the resulting ignition regime strongly depends on the resolution and the use of the burning limiter: most of the successful detonations in the lower-resolution run turned out to be false for the higher resolution, except for the consistent success in the model having the heaviest core and envelope in their simulation grids, 1.1 M$_{\rm{\odot}}$ and 0.05 M$_{\rm{\odot}}$ respectively, with the mixed composition of $^4$He:$^{12}$C:$^{16}$O=0.6:0.2:0.2. 

In fact, studies on the requirement of the resolution and the criteria of the initiation and quenching of the detonation are relatively matured in the research field of terrestrial detonation. The terrestrial detonation theories for describing them are supported by experimental evidence at hand, and it has been common for several decades to apply such theories to design detonation engines and detonation-quenching devices.

\begin{figure}[h]
\centering
\scalebox{0.6}{\includegraphics[trim={90 12 120 3},clip]{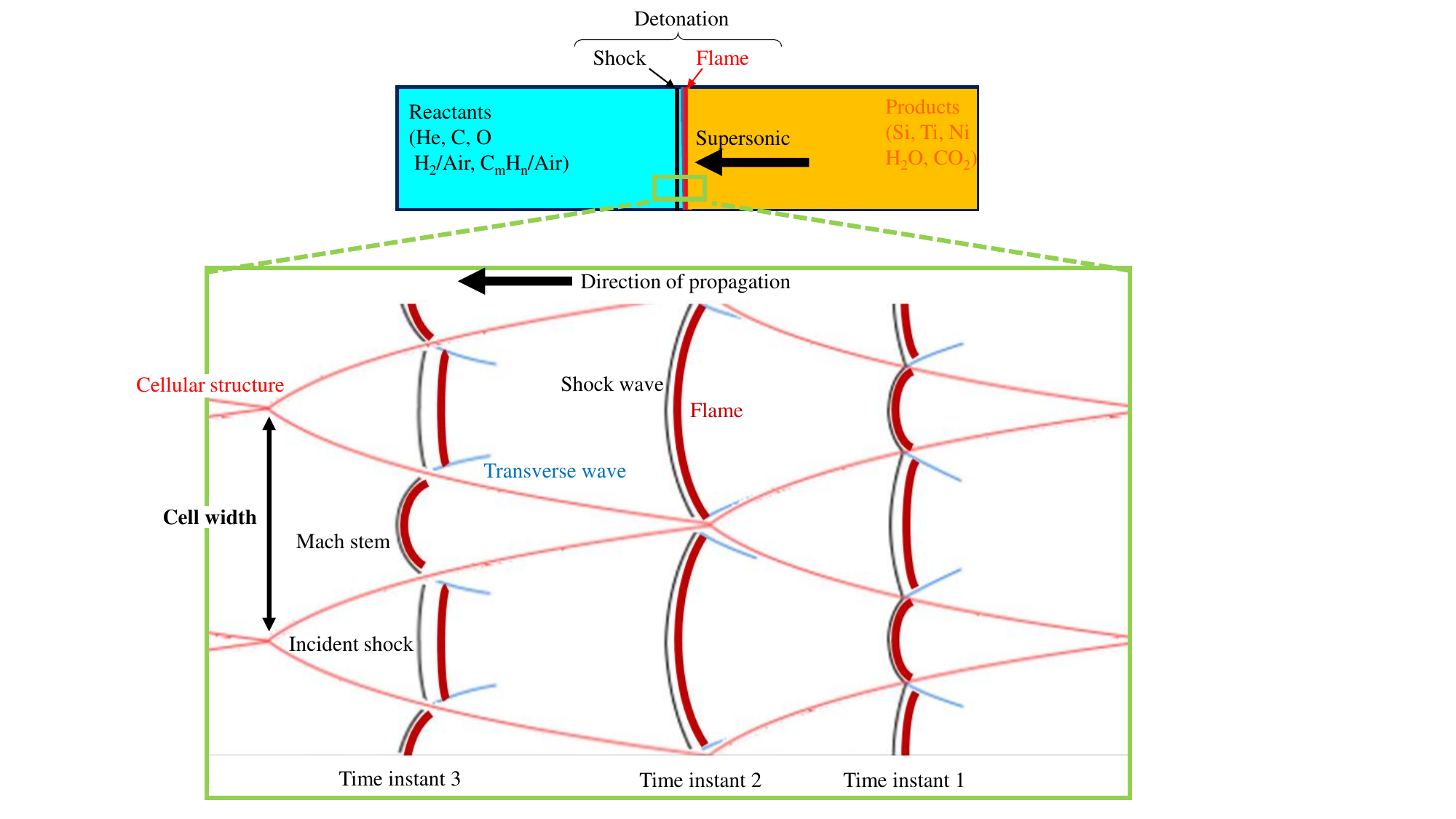}}
\caption{Schematic of cellular structure of detonation.  \label{fig:schematic}}
\end{figure}

Most of the detonation theories are based on the cellular structure, which expresses a multi-dimensional unsteady shock-flame complex schematically illustrated in Fig. \ref{fig:schematic}; detonation is inherently multi-dimensional. It is unsteady consisting of an alternate pattern of Mach stem and incident shock, which are a stronger overdriven detonation and a weaker shock-flame complex, respectively. Transverse waves propagate along these shock waves, forming a triple point at their intersection. At the moment transverse waves collide with each other, micro explosion occurs driving the next life cycle of the transverse waves.

The cellular structure was originally discovered in terrestrial systems by using a soot-foil as the trace of the locus of the triple point. As several studies on thermonuclear detonation demonstrated \citep{Boisseau_1996,Gamezo_1999,Timmes_2000,Papatheodore_2014}, the cellular structure is indeed common between terrestrial and astrophysical detonations. Cell width, which is defined as the lateral size of the cellular structure, is the most important spatial scale in detonation. It is widely used for providing the criteria of initiation and quenching in the field of terrestrial detonation.  In addition, cell width could be an indicator of the resolution level required in the simulations; the cell should be resolved for robust estimation of the success or failure of detonation. As has been indicated by some terrestrial and astrophysical studies \citep{Tsuboi_2008,Guillochon_2011,Glasner_2018}, numeric initiation can occur with an insufficient resolution if the reaction time scale is much shorter than the sound crossing time of the computational mesh.

Despite its importance in the discussion of the resolution dependence and the criteria for the success or failure, it is not common in studying SN Ia explosions to address the cellular dynamics. In addition to the important two theories of steady detonation, Chapman-Jouguet (C-J) theory \citep{Chapman1899} and Zeldovich-von-Neumann-Doring (ZND) theory that have been frequently used in the context of SNe Ia \citep[e.g.,][]{TimmesNiemeyer_2000,Shen_2014,Moore_2013}, it is worthwhile applying the terrestrial cell-based theories to address the issues of astrophysical detonation. We note that the extreme difference between the two systems (terrestrial and astrophysical detonations) in the orders of density and temperature brings additional factors to consider, such as electron degeneracy, energy and pressure of radiation, and so on. The multi-stage nature of nuclear burning and its extreme temperature sensitivity could also come into play. Nevertheless, incorporating the cell-based theories may form an important step in opening up new methodologies in astrophysics connected to terrestrial laboratory experiments.

In the present work, we focus on the helium detonation initiated in the helium-rich envelope of a WD, which is the first stage of the double detonation model. The cellular structure has not been investigated in such systems except for few studies, such as \cite{Moore_2013} in which the cellular structure was demonstrated on the envelope base of WD. \cite{IwataMaeda2024} also showed the development of the cellular structure in the helium detonation, and further presented the first trial to address the threshold of initiation and quenching of detonation via the cell-based theories and experimental laws of terrestrial detonation, with a success in reproducing the trends found in previous hydrodynamics results of \cite{Shen_2014} and \cite{Moore_2013}.

As an extension of \cite{IwataMaeda2024}, which focused on the application of the observed cell sizes to predict the general outcomes of the helium detonation, we address full details of the helium-detonation physics in the present study, including  ZND theoretical features, unsteady wave motions, chemical abundance structure, and the numerical soot-foil records. An insight into the requirement of resolution for helium-rich detonation in a WD envelope is also provided.

\section{Numerical setup} \label{sec:setup}

\subsection{Code configuration} \label{subsec:code}

Our 2D hydrodynamic code, which was used in \cite{IwataMaeda2024} is an extension of 1D version described in \cite{Iwata_2022}, except for the absence of gravity because the present study focuses on the local detonation structure in a uniform medium. This code was originally developed for solving terrestrial detonation problems \citep{Iwata_2016, Iwata_2022, Iwata_2023}. Its chemical reaction solver has been replaced by the one for the nuclear reaction network of 13 isotopes. It solves Euler equation systems, closed by an equation of state (EOS) for high-density fully ionized plasma \citep{Timmes_2000}. 

\cite{Shen_2014} and several subsequent studies suggested that relatively minor isotopes such as $^{14}\rm{N}$ formed by CNO cycles; protons ejected through $^{14}\rm{N}(\alpha,\gamma)^{18}\rm{F}(\alpha,\it{p})^{\rm{21}}\rm{Ne}$ can boost the formation of $^{16}\rm{O}$ through the proton-catalyzed reaction sequence of $^{12}\rm{C}(\it{p},\gamma)^{\rm{13}}\rm{N}(\alpha,\it{p})^{\rm{16}}\rm{O}$. We defer including this aspect to our future work, since our main focus is the comparison with previous studies that adopted the same 13-isotope network. Also, the smaller network benefits to exploring a wide parameter space with a reasonable computational cost. 

Discretization schemes are common to our previous studies \citep{Iwata_2023,Iwata_2022}. Advection fluxes are calculated by using the AUSM+up scheme with the minmod limiter \citep{KitamuraShima2013}, which can apply to flows with a range of Mach number from low subsonic to hypersonic ones and has been used for a variety of flow problems in aerospace engineering. An operator splitting method is used to evolve the flow and reaction separately in time: an explicit time integration for the flow by third-order total-variation-diminishing Runge-Kutta scheme \citep{ShuOsher1988} and an iterative implicit time integration for the reaction by the variable-order Bader-Deuflhard method \citep{BaderDeuflhard_1983}.

\subsection{Model setup} \label{subsec:model}

Our 2D model setup follows that of \cite{Gamezo_1999}. A rectangular area is discretized to 2000$\times$1200 uniform square Eulerian meshes. The adaptive mesh refinement technique is not used in the present work. A detonation-fixed coordinate is realized by adding inertial force to control the longitudinal position of detonation. For this setup, the fuel He/C/O mixture enters the computational domain from the left boundary at detonation velocity, and the density and composition of the incoming flow are treated as parameters. This model configuration is commonly used in terrestrial detonation problems to investigate the C-J detonation, which is defined as the steady detonation around the C-J theoretical velocity. Cell width of the C-J detonation is used as a basic quantity in forming the detonation criteria in terrestrial systems (see Detonation database by \cite{Detonation_database}). 

Periodic condition is imposed on the lateral (upper and lower) boundaries, unlike \cite{Gamezo_1999} who applied reflective boundaries. \cite{Boisseau_1996} indicated that the influence of the boundary treatment is negligible as far as the computational domain is sufficiently large to capture the cellular structures as in the present work. One-dimensional ZND profiles are pasted on the rectangular domain as an initial condition, disturbed either periodically by placing cold spots behind the shock front or randomly by displacing the shock at each lateral position following a white noise pattern created in advance. The choice of the methods to disturb the ZND profiles depends on the computational cases; a white-noise perturbation, which gives randomly arranged amplitudes with a uniform spectrum, is applied to the more helium-rich detonation, which needs more strongly disturbed initial fronts to initiate the cellular motion. Actually, the observed cellular structure did not depend on the choice of the initial disturbance. On the outlet (right) boundary, primary variables are fixed to be those from the ZND theory at the corresponding post-shock distance to keep the C-J state of the detonations with a thermodynamic consistency. Toward the outlet boundary, 400 meshes in the x-direction are extended logarithmically, so that the spurious disturbance coming from this fixed outlet boundary is minimized.

Table \ref{tab:parameter} summarizes a parameter space in this study along with characteristic reaction scales and the resulting cell widths. The upstream density is varied in the range of $10^5-10^6$ g cm$^{-3}$, which covers the condition at the base of a WD envelope with core masses of 0.9-1.1 M$_\odot$ and envelope masses of 0.01-0.05 M$_\odot$. Mass fraction of $^{\rm{4}}{\rm{He}}$ $X_{\rm{He}}$ is taken to be between 0.6-1.0 or 0.0, and the remaining fraction is divided equally in masses to $^{\rm{12}}{\rm{C}}$ and $^{\rm{16}}{\rm{O}}$. The choice of the upstream temperature $T_{\rm{1}}$ was confirmed not to affect the structure of the detonations, and we fix it as $T_1$=$2{\times}10^8$ K for all the models. 

\begin{deluxetable*}{ccccccccc}
\tablenum{1}
\tablecaption{Reaction length scales and cell width of each computational case. Parenthesized values are shown as the length at which the isotope is half consumed after reaching the maximum. \label{tab:parameter}}
\tablewidth{0pt}
\tablehead{
\colhead{$\it{X}_{\rm{He}}$} & \colhead{Density} & \colhead{$\it{L}_{\rm{\dot{\varepsilon}max}}$} & 
\colhead{$\it{L}_{\rm{q}}$} &
\colhead{$\it{L}_{\rm{He}}$} & \colhead{$\it{L}_{\rm{C}}$} & \colhead{$\it{L}_{\rm{O}}$} & \colhead{$\it{\lambda}$}   \\
\colhead{ } & \colhead{$10^5$g cm$^{-3}$} & \colhead{cm} & \colhead{cm} &
\colhead{cm} & \colhead{cm} & \colhead{cm} & \colhead{cm}
}
\startdata
0.0 & 1 & $6.9\times10^5$ & $1.1\times10^7$ & ($
1.8\times10^6$) & $1.3\times10^7$ & $> 1\times10^{13}$ & $8.9\times10^7$ \\
0.0 & 2& $2.7\times10^4$ & $3.9\times10^5$  & ($7.2\times10^4$) & $4.4\times10^5$ & $4.7\times10^{12}$ & $2.4\times10^6$ \\
0.0 & 5& $1.5\times10^3$ & $6.1\times10^3$  & ($2.9\times10^3$) & $7.0\times10^3$ & $6.5\times10^9$ & $4.0\times10^4$ \\
0.0 & 10& $1.9\times10^2$ & $4.0\times10^2$  & ($3.8\times10^2$) & $3.9\times10^2$ & $5.0\times10^7$ & $3.0\times10^3$ \\
0.6 & 1 & $1.0\times10^3$ & $3.5\times10^3$  & $1.3\times10^5$ & $2.8\times10^6$ & $2.3\times10^3$ & $2.5\times10^4$ \\
0.6 & 2 & $2.6\times10^2$ & $9.6\times10^2$  & $1.4\times10^4$ & $1.8\times10^5$ & $7.7\times10^2$ & $1.0\times10^4$ \\
0.6 & 5 & $5.2\times10^1$ & $1.8\times10^2$  & $1.3\times10^3$ & $3.0\times10^3$ & $2.2\times10^2$ & $2.2\times10^3$ \\
0.6 & 10 & $3.2\times10^1$ & $4.6\times10^1$ & $2.7\times10^2$ & $1.4\times10^2$ & $1.0\times10^2$ & $8.0\times10^2$  \\
0.8 & 1 & $7.6\times10^2$ & $1.0\times10^4$ & $2.1\times10^6$ & $3.8\times10^5$ & $1.5\times10^3$ & $1.0\times10^5$ \\
0.8 & 2 & $1.9\times10^2$  & $2.0\times10^3$ & $2.3\times10^5$ & $4.4\times10^4$ & $5.1\times10^2$ & $1.7\times10^4$  \\
0.8 & 5 & $4.1\times10^1$ & $3.0\times10^2$ & $1.5\times10^4$ & $2.0\times10^3$ & $1.4\times10^2$ & $2.2\times10^3$ \\
0.8 & 10 & $3.0\times10^1$ & $7.7\times10^1$ & $2.0\times10^3$ & $1.6\times10^2$ & $6.0\times10^1$ & $9.6\times10^2$ \\
0.9 & 1 & $6.6\times10^2$ & $8.7\times10^4$ & $6.0\times10^7$ & $2.1\times10^5$ & $1.3\times10^3$ & $8.0\times10^5$\\
0.9 & 2 & $1.7\times10^2$ & $1.3\times10^4$ & $2.0\times10^7$ & $2.8\times10^4$ & $4.3\times10^2$ & $8.0\times10^4$ \\
0.9 & 5 & $3.8\times10^1$ & $1.1\times10^3$ & $5.2\times10^6$ & $1.7\times10^3$ & $1.2\times10^2$  & $8.0\times10^3$ \\
0.9 & 10 & $2.7\times10^1$ & $2.1\times10^2$ & $1.9\times10^6$ & $2.0\times10^2$ & $5.0\times10^1$  & $1.2\times10^3$ \\
1.0 & 1 & $6.9\times10^5$ & $3.8\times10^6$ & $6.5\times10^7$ & ($6.3\times10^6$) & ($7.5\times10^6$)  & $4.0\times10^7$ \\
1.0 & 2 & $1.6\times10^5$ & $1.1\times10^6$ & $2.1\times10^7$ & ($1.5\times10^6$) & ($1.9\times10^6$)  & $1.0\times10^7$ \\
1.0 & 5 & $1.8\times10^4$ & $2.5\times10^5$ & $5.2\times10^6$ & ($2.9\times10^5$) & ($8.6\times10^5$) & $3.0\times10^6$ \\
1.0 & 10 & $2.7\times10^3$ & $9.2\times10^4$ & $1.8\times10^6$ & ($8.7\times10^4$) & ($1.3\times10^5$) & $1.5\times10^6$ \\
\enddata
\end{deluxetable*}

\section{1D-ZND spatial scales and resolution} \label{sec:znd}

Choosing the resolution for simulating the cellular detonation is one major issue that has been investigated extensively in terrestrial detonation, because it affects the size of the observed cellular structure. It is common to locate a few tens of meshes in the half-reaction length (hrl), which is defined as the distance between the shock and the position where the fuel is half consumed. Recent large-scale simulations use around, or even more than, 60 points/hrl. \cite{Mazaheri2013} proposed that the half-reaction length has to be resolved by not less than 24 meshes for a relatively 'regular' mixture, in which the post-shock reaction rate is not sensitive to the shock-speed variation and thereby exhibits regular/uniform cellular patterns. 

Complication in choosing the resolution in astrophysical detonation problems originates from the fact that scales for heat release, temperature increase, and fuel consumption could differ by orders of magnitude. In addition, the multi-stage nature of the reaction sequences ($^{\rm{4}}$He to $^{\rm{12}}$C, $^{\rm{12}}$C to $^{\rm{16}}$O, $^{\rm{16}}$O to $^{\rm{28}}$Si, and $^{\rm{28}}$Si to $^{\rm{56}}$Ni) further complicates this issue. This is not a problem in terrestrial detonation, where the spatial scale for the heat release, the temperature increase, and half-reaction length of each reactant are all close. \cite{Timmes_2000} concluded that 80 points per burning length scale, which is defined by the specific internal energy and the post-shock heat release, are sufficient to converge in the predicted cell size of $^{\rm{12}}$C burning. \cite{Gamezo_1999} chose their uniform mesh sizes based on the half-reaction lengths of $^{\rm{12}}$C, $^{\rm{16}}$O, and $^{\rm{28}}$Si and discussed the cell size for each reaction stage separately.

The ZND theory, which describes 1D theoretical structure of steady detonation, is used to quantify the spatial scales in helium-rich detonation to choose the resolution in the present 2D studies. The 1D governing equation system shown below considers a steady reacting flow, neglecting viscous and lateral fluxes:

\begin{eqnarray}\label{eq:znd}
\frac{d}{dx}(\rho u) &=& 0 \ ,\\
\frac{d}{dx}(\rho u^2+P) &=& 0 \ ,\\
\frac{d}{dx}(h+\frac{u^2}{2}-q) &=& 0 \ , {\rm and}\\
\frac{d}{dx}(\rho uX_i) &=& \rho W_i \ ,
\end{eqnarray}
where $h, q, X_i,$ and $W_i$ are enthalpy, accumulated nuclear energy generation, mass fraction of each isotope, and production rate of each isotope by nuclear reactions, respectively. Starting from the Neumann-spike state calculated just behind the C-J speed shock front, the equations are integrated along with the 13-isotope reaction network until the sonic condition is reached.

\subsection{1D-ZND profiles} \label{subsec:1dznd}

\begin{figure}[h]
\centering
\scalebox{0.9}{\includegraphics[trim={250 0 20 0},clip]{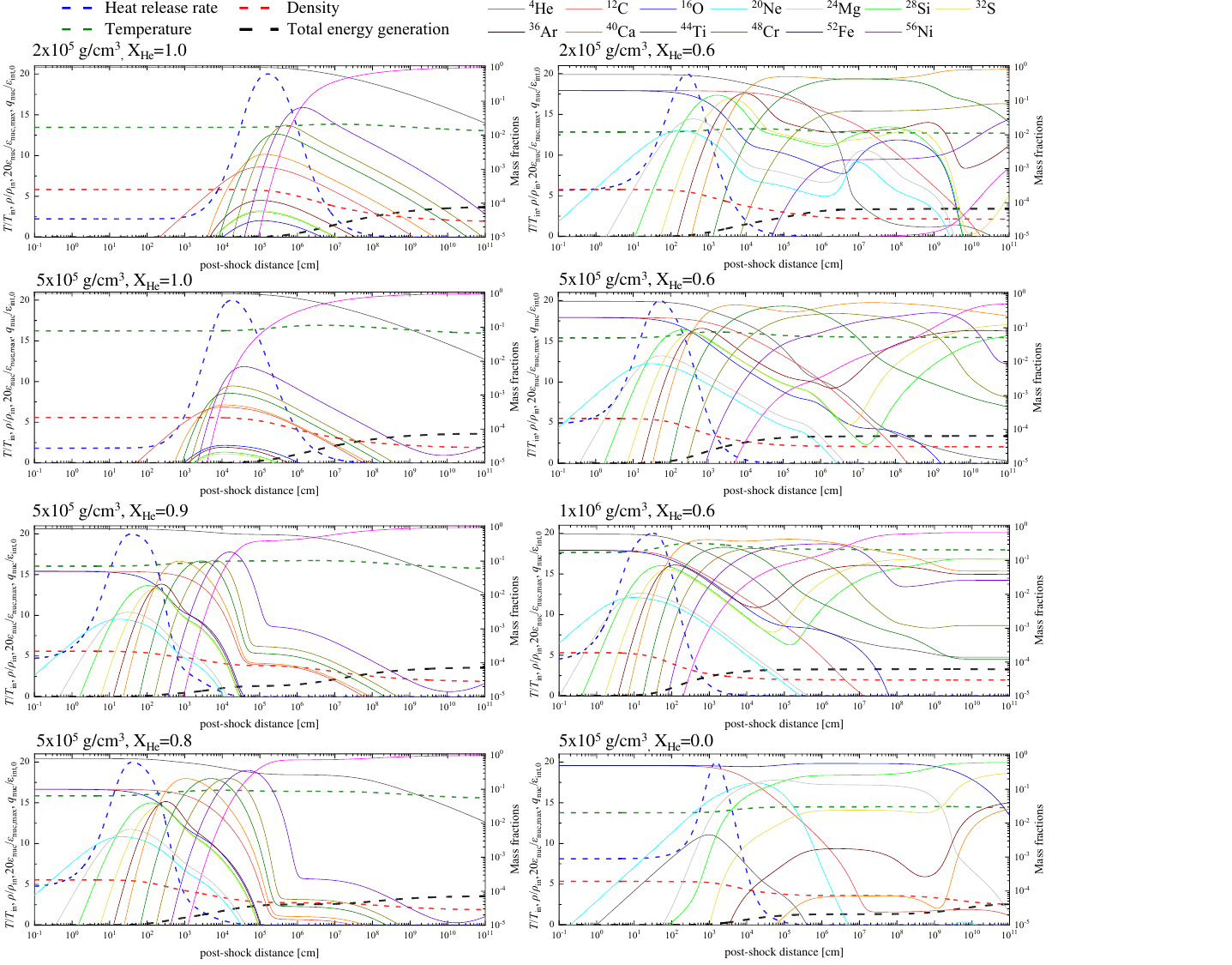}}
\caption{1D ZND profiles of the nuclear energy release rate, density, temperature, total energy generation (left vertical axes), and mass fractions of each isotope (right vertical axes). Eight cases with different combinations of density and initial helium mass fraction are shown here. The vertical axes are all shown in logarithmic scale. The abscissa is the post-shock distance shown in logarithmic scale. \label{fig:znd}}
\end{figure}

Fig. \ref{fig:znd} shows 1D ZND profiles of eight selected cases presented in Table \ref{tab:parameter}. Temperature and density are shown as normalized by the upstream value, and nuclear energy generation rate $\dot{\varepsilon}_{\rm{nuc}}$ are normalized by the maximum values. The total accumulated amount of the energy generation $q$ is normalized by the post-shock internal energy ${\epsilon}_{\rm{int,0}}$. 

For the pure-helium conditions illustrated in two panels of the left column of Fig. \ref{fig:znd}, the general reaction process is relatively simple, where the maximum heat release associated with the production of $^{\rm{44}}$Ti, $^{\rm{48}}$Cr and $^{\rm{52}}$Fe is followed by the consumption of $^{\rm{4}}$He and the production of $^{\rm{56}}$Ni. Comparing the ZND profiles of different densities, it is seen that they are qualitatively similar and that the reaction progresses faster for the higher density. This is similarly true for another helium mass fraction of 0.6 (right panels), except that the difference in the consumption scale between $^{\rm{12}}$C and $^{\rm{16}}$O diminishes as the density increases.

When $X_{\rm{He}}$=0.6-0.8, the presence of $^{\rm{12}}$C and $^{\rm{16}}$O complicates the reaction progress. The maximum heat release associated with the consumption of $^{\rm{16}}$O and the production of IMEs is followed by sequential formation of heavier isotopes, in which $^{\rm{4}}$He and $^{\rm{12}}$C are consumed gradually. Also, $^{\rm{20}}$Ne starts to be produced soon behind the shock due to the $\alpha$-capture of $^{\rm{16}}$O, as it is the case for $X_{\rm{He}}$=0.0. The cases $X_{\rm{He}}$=0.8 and 0.9 look similar in the reaction progress, in which $^{\rm{56}}$Ni becomes the major component after consuming almost all the lighter isotopes. Further increase of $^{\rm{56}}$Ni toward the NSE occurs through the second heat generation coincident with the accelerated decrease of $^{\rm{4}}$He and nearly the completed consumption of $^{\rm{12}}$C and $^{\rm{16}}$O. The reaction sequences are much more complicated for the $X_{\rm{He}}$=0.6 cases, in which several IMEs including $^{\rm{28}}$Si, $^{\rm{32}}$S, $^{\rm{36}}$Ar, and $^{\rm{40}}$Ca turn to increase after the consumption stage. In these cases, the formation of $^{\rm{56}}$Ni is delayed and does not immediately follow the consumption of helium, but occurs in the larger scales after the plateau of IMEs such as $^{\rm{40}}$Ca and $^{\rm{44}}$Ti. The explanation for this is given by the amount of $^{\rm{4}}$He, $^{\rm{12}}$C and $^{\rm{16}}$O at each reaction stage: the formation of heavier isotopes is halted once $^{\rm{4}}$He decreases to be $\sim 10^{-1}$ in the mass fraction, since the pre-existing helium may not be sufficiently abundant to bring all the heavy elements into the NSE/QSE. It then has to wait for the production of helium through heavy-ion reactions such as $^{12}$C+$^{12}$C and $^{12}$C+$^{16}$O. The additional helium thus created completes the reaction sequences, e.g., reaching the production of IGEs. A caveat is that this behavior of the reaction process may depend on the choice of the reaction network, for which a more detailed network should be considered; we will postpone such investigation to the future.

In the carbon-oxygen media ($X_{\rm{He}}$=0.0), the formation of IGEs is extremely delayed, ending up with the formation of IMEs in the WD scale. Also, the consumption scale of $^{\rm{12}}$C and $^{\rm{16}}$O is much different, highlighting the multi-stage nature of the reactions. These behaviors can be explained by the slow injection of helium by heavy-ion reactions. Nevertheless, unlike the higher densities explored by \cite{Gamezo_1999}, nuclear energy generation is concentrated in one location; multiple heat release scales are less evident. It is also common to every case with pre-existing helium that the peak of nuclear energy generation and associated temperature increase occur once almost at the same time. Notable increase in the accumulated energy generation occurs twice in some cases, but the second increase is not linked to the temperature change. 

Based on the 1D theoretical structures described above, some characteristic spatial scales are computed as shown in Table \ref{tab:parameter}. ${\it{L}}_{\rm{\dot{\varepsilon}}_{\rm{max}}}$ is the distance at which the instantaneous energy generation reaches the maximum. ${\it{L}}_{\rm{q}}$ is the distance at which $\it{q}$ reaches a quarter of $\epsilon_{\rm{int,0}}$, which is the best coincident with the temperature increase throughout the parameter space. $\it{L}_{\rm{He}}$, $\it{L}_{\rm{C}}$, and $\it{L}_{\rm{O}}$ are the half-reaction length of each isotope in the initial mixture; for this definition, the length values for the components that are absent in the initial mixture are parenthesized, computed as the position at which the mass fraction reduces to half the maximum value.

\subsection{Choosing resolution} \label{subsec:1dznd}

Then, how can we define the characteristic spatial scale that must be resolved for these complicated multi-scale detonations? Basically, the motion of transverse waves is driven primarily by the inherent instability of the shock front with exothermic reaction \citep{lee_2008}; when the shock front becomes stronger, the subsequent reaction is intensified, further accelerating the shock front with the help of compression waves. When the shock front becomes weaker, the opposite occurs. Therefore, the spatial scale for exothermicity, or the temperature increase, serves as a fundamental quantity. 

In terrestrial detonation, the scales for the maximum heat release and the half-reaction length are both close to the scale for the temperature increase. The half-reaction length is then customarily used as the criterion of the resolution. As stated previously, the second increase in the total energy generation is not associated with a notable temperature increase. Therefore, the first increase of the total energy generation should be more important as the resolution criterion. The temperature increase is often taken as the resolution indicator in terrestrial detonation, where the temperature increases by an order, whereas the magnitude of the temperature increase in the present study is too mild (by a factor $\sim$ 1.1) to define its scale; this is due to substantial contribution of radiation energy. Hence, we find that $L_{\rm{q}}$ is better as the basic scale to be resolved, which is close to $L_{\rm{C}}$ for $X_{\rm{He}}$=0.0 cases, and $L_{\rm{O}}$ for $X_{\rm{He}}$=0.6,0.9,1.0 cases. In the following 2D study, twenty meshes are placed within $L_{\rm{q}}$. This resolution is also sufficient to resolve the burning length scales by more than 80 meshes in all the present models.

\subsection{Check of convergence and consistency} \label{subsec:1dznd}

The resolution dependence of our solution was checked by changing the mesh sizes; we tested 1, 2, 5, and 7 cm in the 2D simulations with $X_{\rm{He}}= 0.60$ and $10^6$ $\rm{g} \ \rm{cm}^{-3}$. The cell width was found to be 8.0$\times10^2$ cm for the standard mesh size of 2 cm, and varied to 6.0, 9.3, and 6.3 $\times 10^2$ cm for 1, 5, and 7 cm resolution, respectively. Thus, the uncertainty was within a factor of 1.6 compared to the standard resolution. Considering that even the most recent hydrodynamic studies on terrestrial detonation suffer from the resolution-dependent cell widths that fluctuate by a factor of a few \citep[e.g., see][]{CRANE20232915}, this is acceptable to judge that the solution has converged. The non-monotonic resolution dependence observed here is also common in terrestrial studies, for which the trend is hard to predict. In addition, as clarified by \cite{IwataMaeda2024}, this level of uncertainty in the cell width has only a minor influence in analyzing the criteria for the initiation and quenching of detonation.

\begin{figure}[h]
\centering
\scalebox{0.58}{\includegraphics[trim={20 150 20 50},clip]{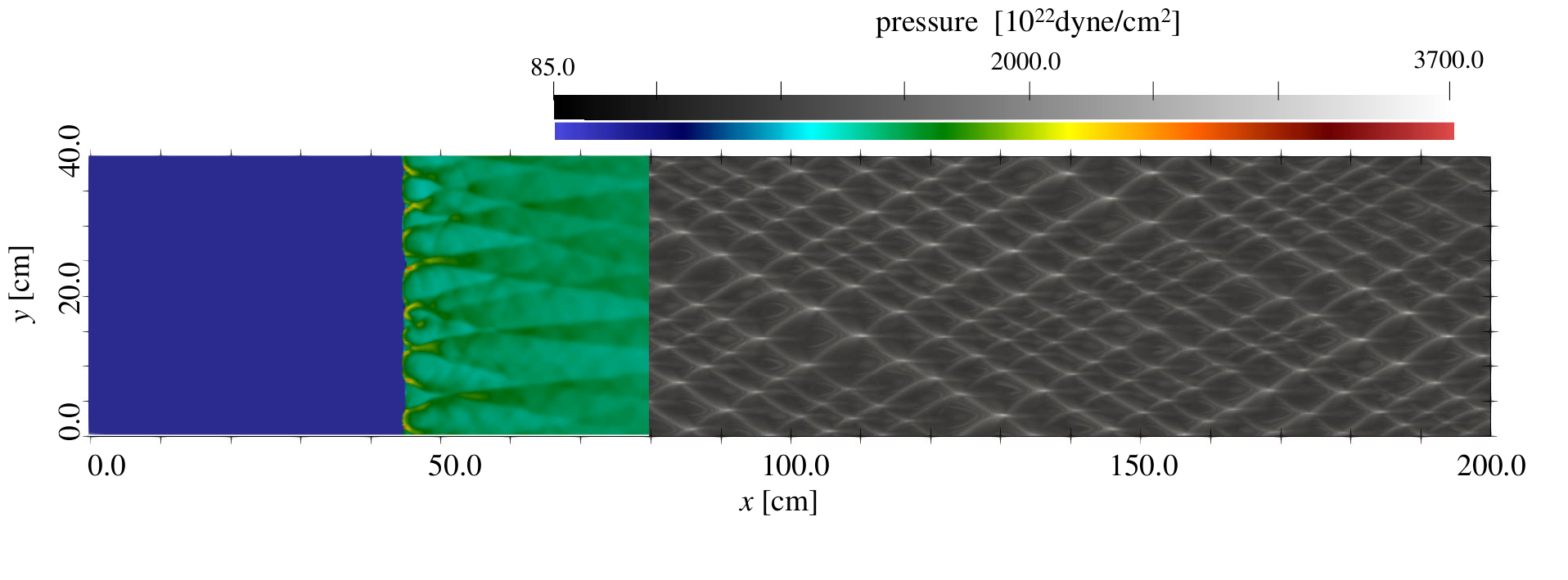}}
\scalebox{0.58}{\includegraphics[trim={20 150 20 50},clip]{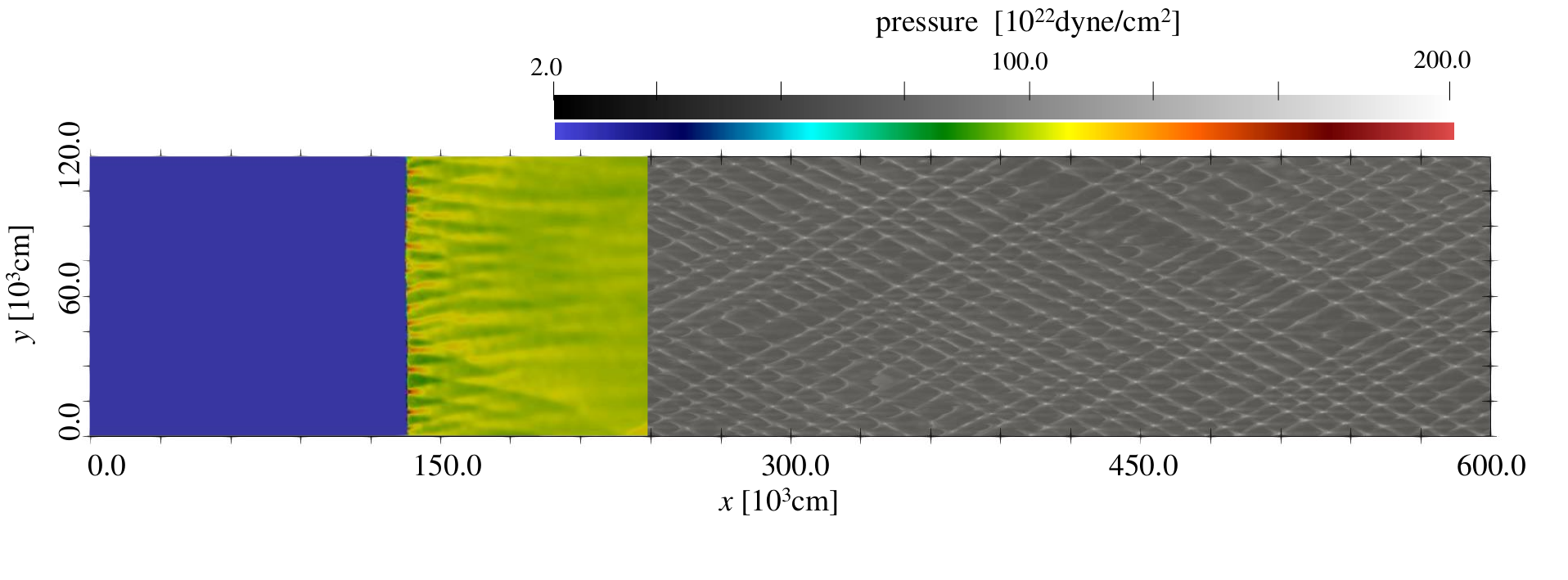}}
\caption{Maximum pressure histories (gray) and overlaid instantaneous pressure distributions (color) of 2D cellular detonations in $^{12}$C and $^{12}$C/$^{16}$O mixtures computed in previous studies by \cite{Timmes_2000} and \cite{Gamezo_1999}: (upper) $^{12}$C, $10^7$ g/cm$^3$ with $\Delta= 10^{-1}$ cm computed by \cite{Timmes_2000}, (lower) $^{12}$C:$^{16}$O=0.5:0.5, $10^6$ g/cm$^3$ with $\Delta = 3\times10^2$ cm computed by \cite{Gamezo_1999}.   \label{fig:timmes}}
\end{figure}

As a sanity check of the consistency with previous results of the cellular detonation, the same conditions as \cite{Timmes_2000} and \cite{Gamezo_1999} are simulated. Figure \ref{fig:timmes} shows the maximum pressure histories seen in these simulations. Here, the maximum pressure experienced at each mesh is recorded in the laboratory coordinate (moving downstream at the detonation velocity). They act as numerical soot-foil records to trace the loci of the triple points, thereby illustrating the cellular structure. The upstream 20 percent of each figure is overlaid by the instantaneous pressure profile of each case. The upper figure is the result for the pure $^{12}$C with $10^7$ g cm$^{-3}$ addressed by \cite{Timmes_2000}, while the lower figure is for the mixture of equal masses of $^{12}$C and $^{16}$O with $10^6$ g cm$^{-3}$ addressed by \cite{Gamezo_1999}. The mesh sizes are set the same as those in their studies; $10^{-1}$ cm and $3\times10^2$ cm, respectively. For a clearer view of the cellular structure, the lowest quarter part in the lateral direction is shown. The diamond-shaped cellular pattern appears in both cases. 

We note that there are irregular cell sizes, where larger cell widths are observed in some places. This feature is common to terrestrial detonation, and it is induced by the failure in a micro explosion at the collision of transverse waves or by the obscured pressure peak of weaker transverse waves. The `cell regularity' is associated with this irregular shape of the cellular structure, and is related to the activation energy of the reaction in terrestrial detonation. As the activation energy increases, the mixture becomes more sensitive to the variation of the shock speed. Therefore, the success or failure of a micro explosion at the transverse-wave collision is more stochastic for larger activation energy. This induces stronger irregularity in the cellular pattern.

The averaged cell width, as calculated by the number of transverse waves in the lateral direction, in these two simulations are 5.5 cm and 1.2$\times10^4$ cm, respectively. In the previous studies, they were found to be 3.8 cm and 2.0$\times10^4$ cm, respectively. The values agree within a factor of two. The minor differences here can be attributed to differences in the numerical configurations, such as discretization schemes, adaptive mesh refinement, and the treatment of boundary conditions. Thus, we conclude that our methodology provides a reasonable estimate for quantifying the cellular dynamics of astrophysical detonation.

\section{2D Cellular structure} \label{sec:2D}

\subsection{Pressure profiles and numerical soot-foils} \label{subsec:cellpress}

\begin{figure}[h]
\centering
\scalebox{0.54}{\includegraphics[trim={15 230 5 12},clip]{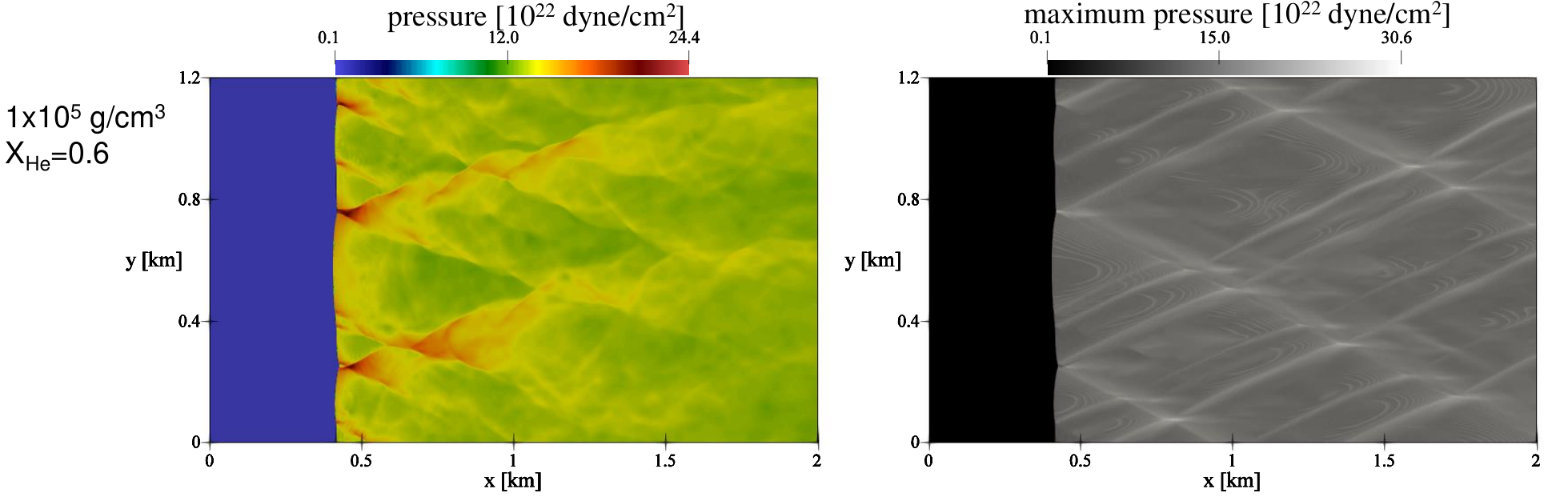}}
\scalebox{0.54}{\includegraphics[trim={15 230 5 12},clip]{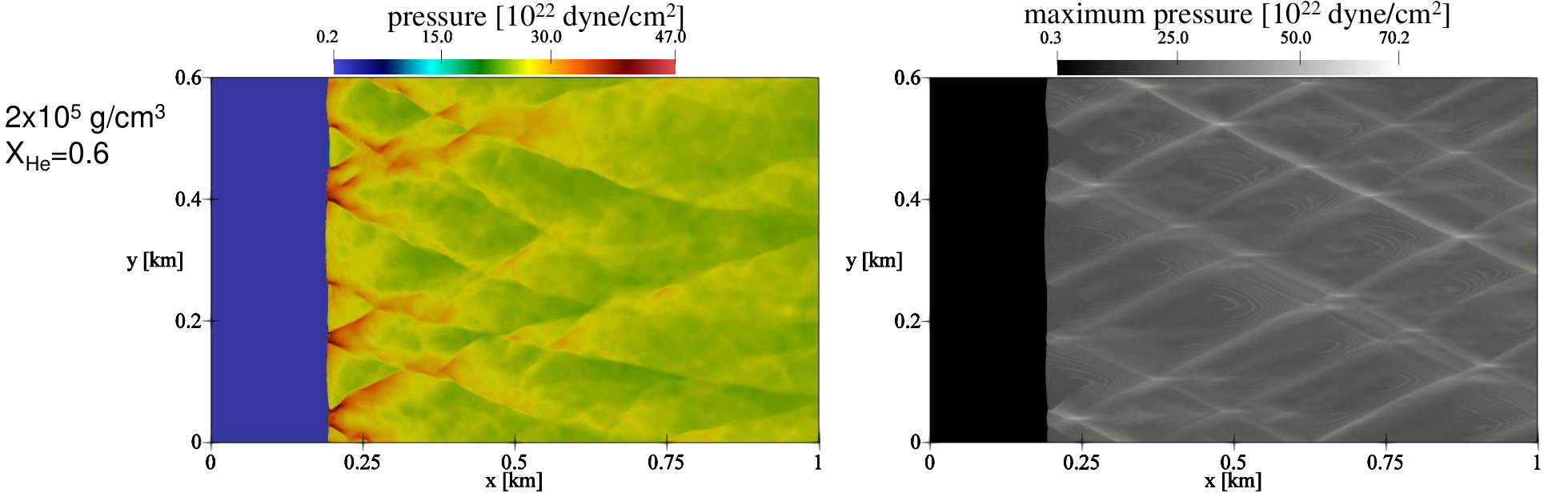}}
\scalebox{0.54}{\includegraphics[trim={15 230 5 12},clip]{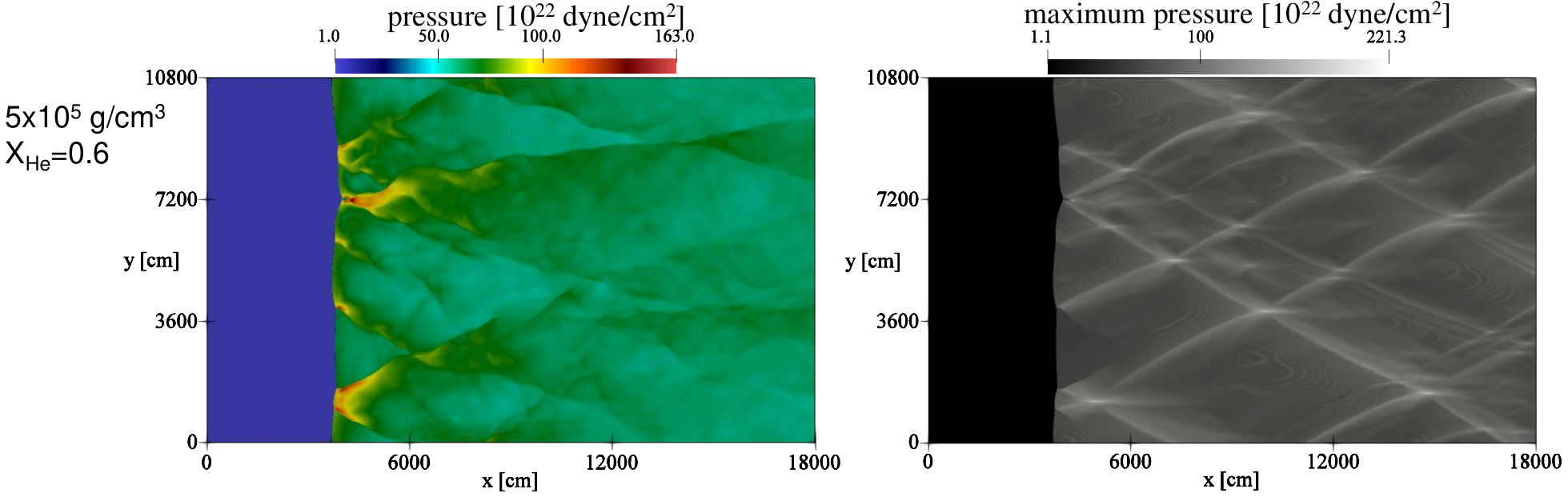}}
\scalebox{0.54}{\includegraphics[trim={15 230 5 12},clip]{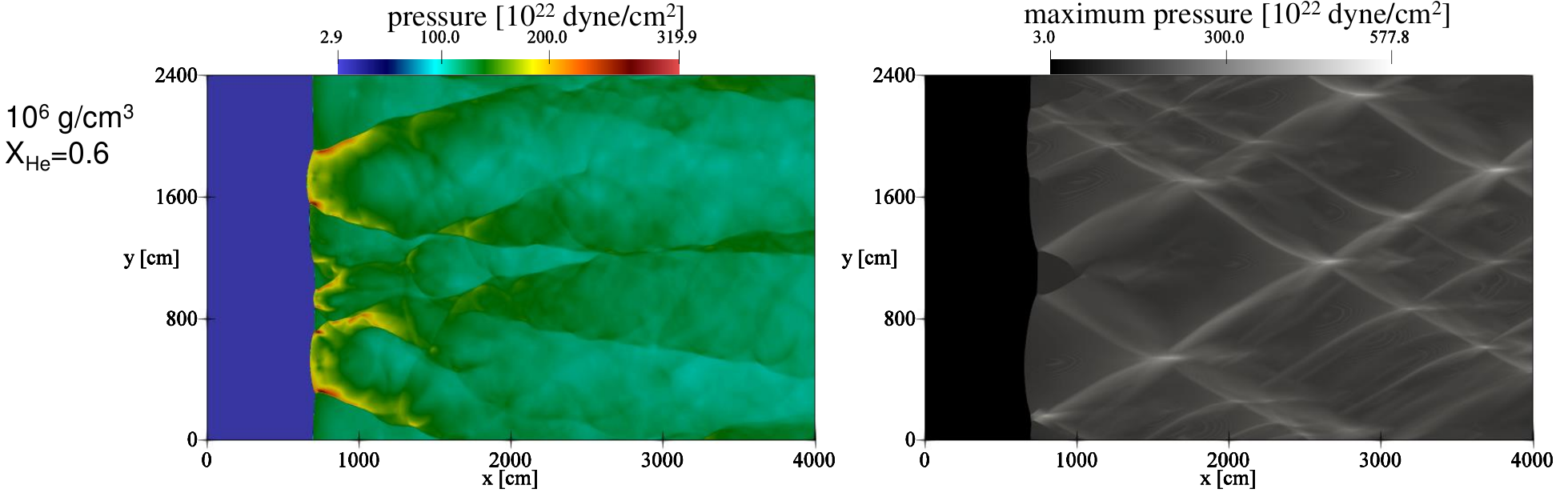}}
\caption{Pressure color maps (the left column) and maximum pressure histories (the right column) of $X_{\rm{He}}$=0.6 He-rich detonations at the envelope of HeCO WD.  \label{fig:pressuremaxp}}
\end{figure}

\begin{figure}[h]
\centering
\scalebox{0.54}{\includegraphics[trim={15 230 5 12},clip]{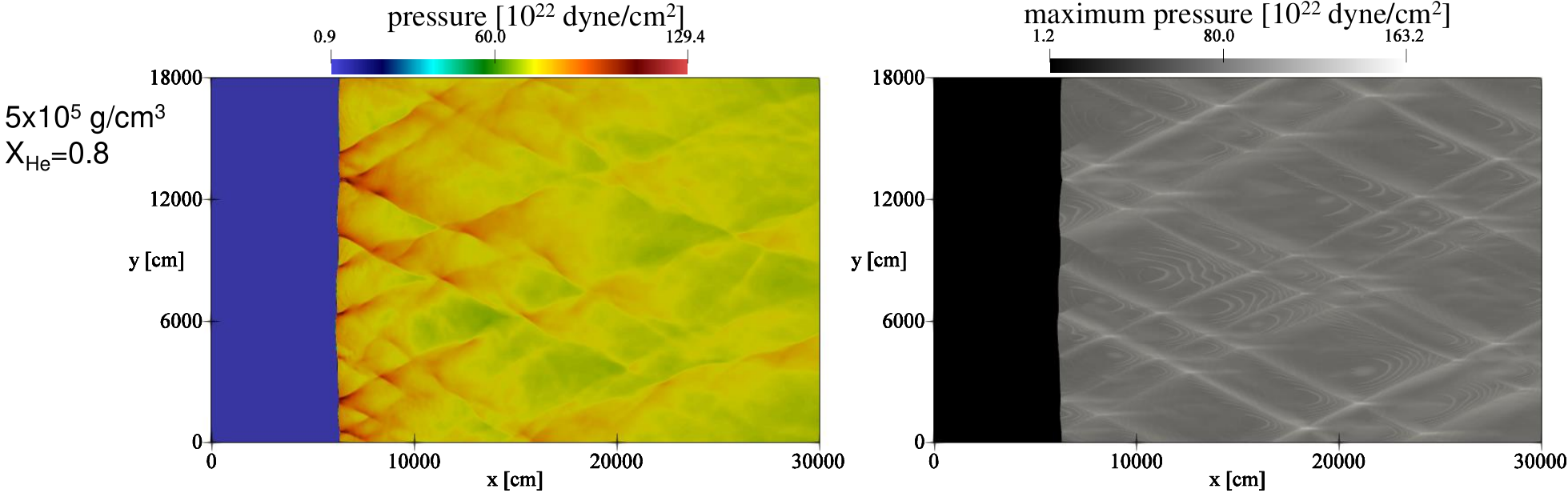}}
\scalebox{0.54}{\includegraphics[trim={15 230 5 12},clip]{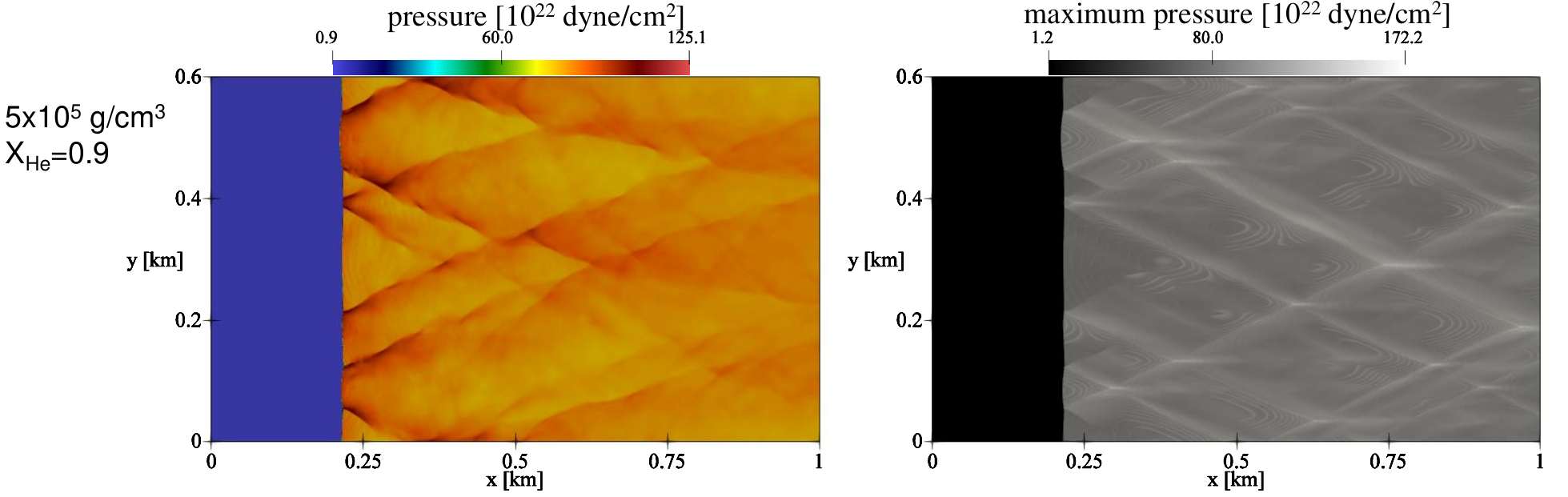}}
\scalebox{0.54}{\includegraphics[trim={15 230 5 12},clip]{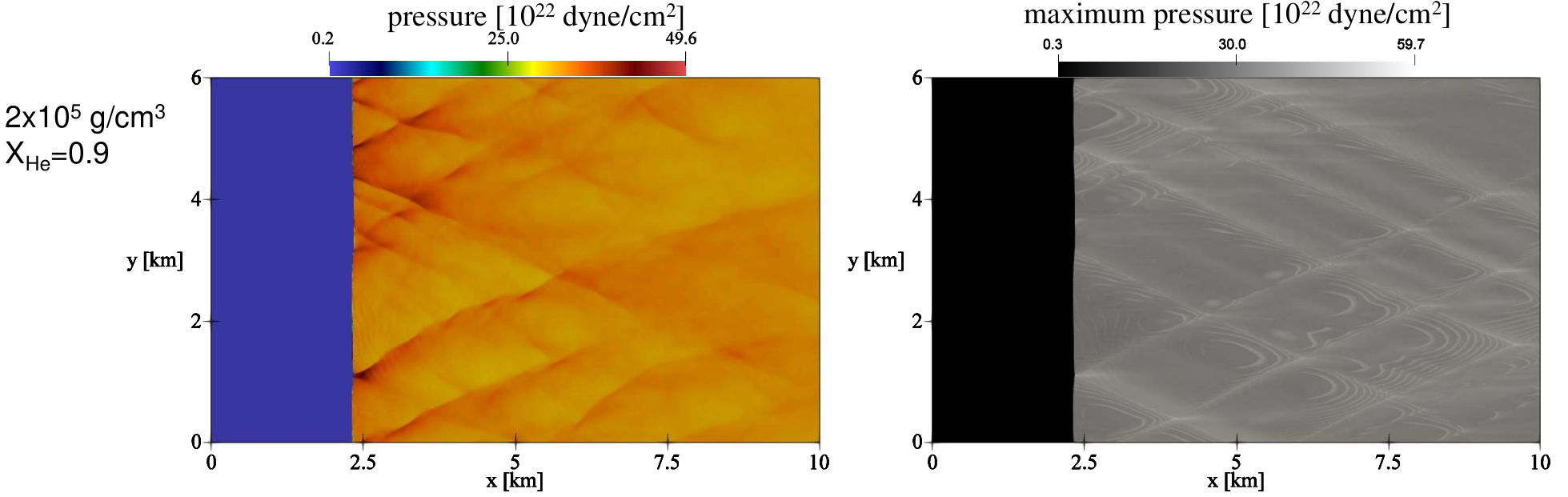}}
\scalebox{0.54}{\includegraphics[trim={15 230 5 12},clip]{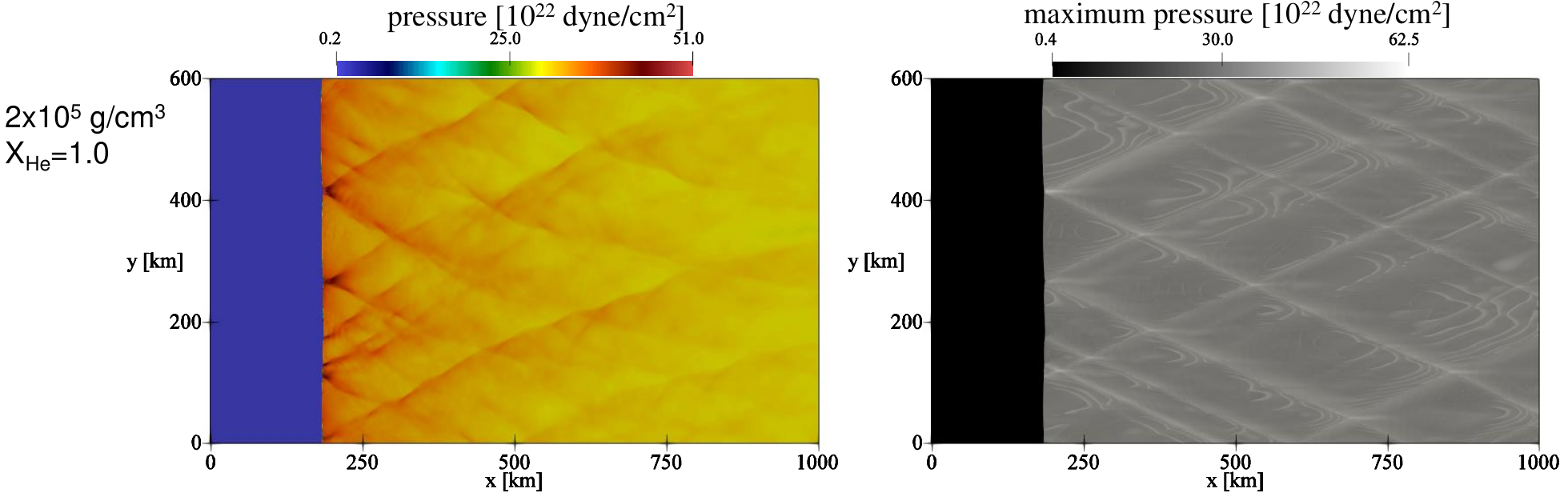}}
\caption{Pressure color maps (the left column) and maximum pressure histories (the right column) of $X_{\rm{He}}$=0.8-1.0 He-rich detonation at the envelope of HeCO WD.  \label{fig:pressuremaxp2}}
\end{figure}

Highly-revolved 2D simulations are performed with the mesh size of $L_q$/20 for each case presented in Table \ref{tab:parameter}. The cellular structure is observed in all the models within our parameter space. Figure \ref{fig:pressuremaxp} (left column) shows snapshots of the pressure fields of four cases adopting $X_{\rm{He}}$=0.60 with the initial densities of 1, 2, 5, and 10 $\times10^5$ g cm$^{-3}$. The corresponding maximum pressure histories are shown in the right column. Four selected cases with $X_{\rm{He}}$=0.80-1.00 are shown in Fig. \ref{fig:pressuremaxp2}. 

\begin{figure}[h]
\centering
\scalebox{0.62}{\includegraphics[trim={27 120 10 130},clip]{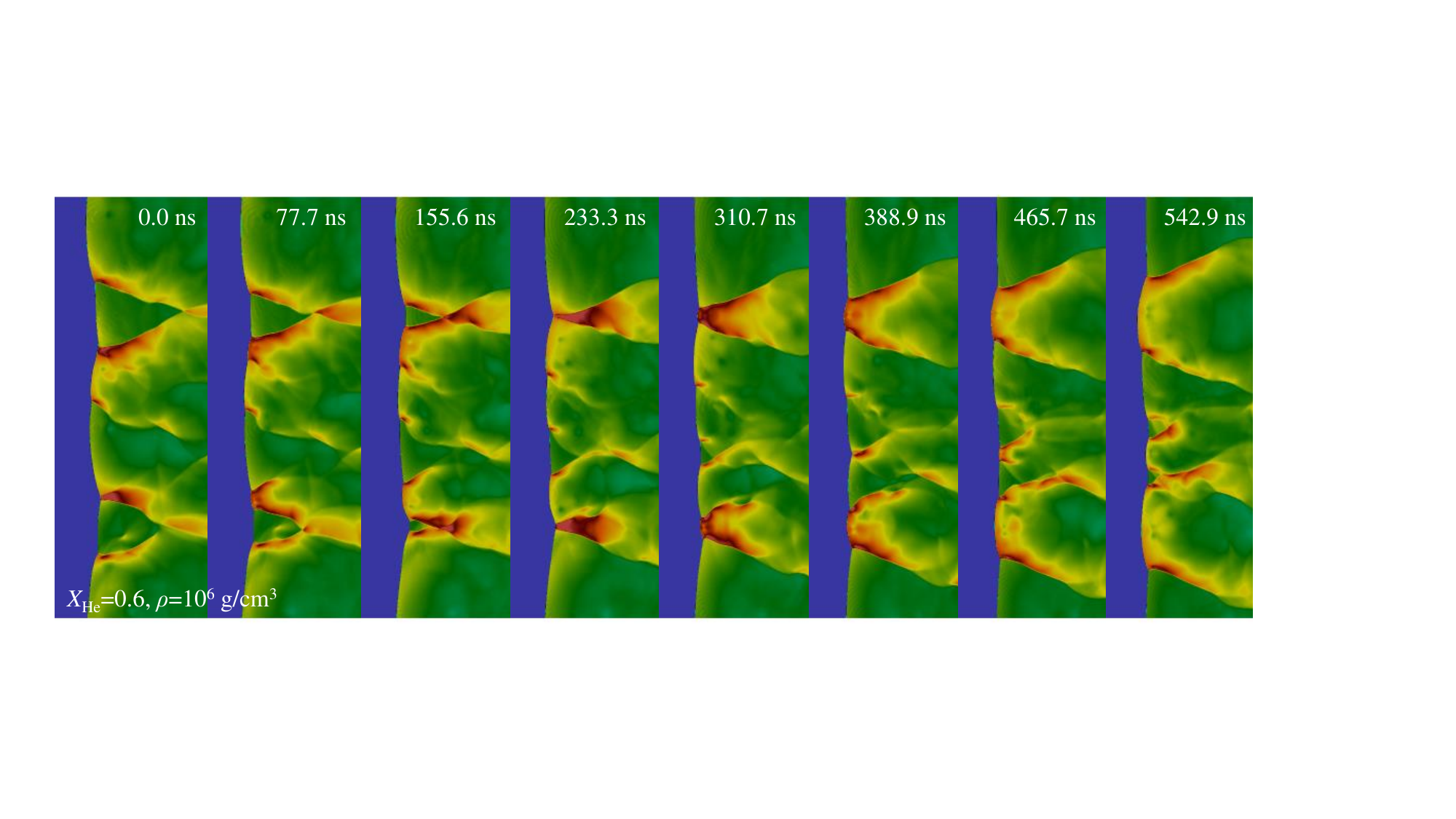}}
\scalebox{0.625}{\includegraphics[trim={33 120 10 130},clip]{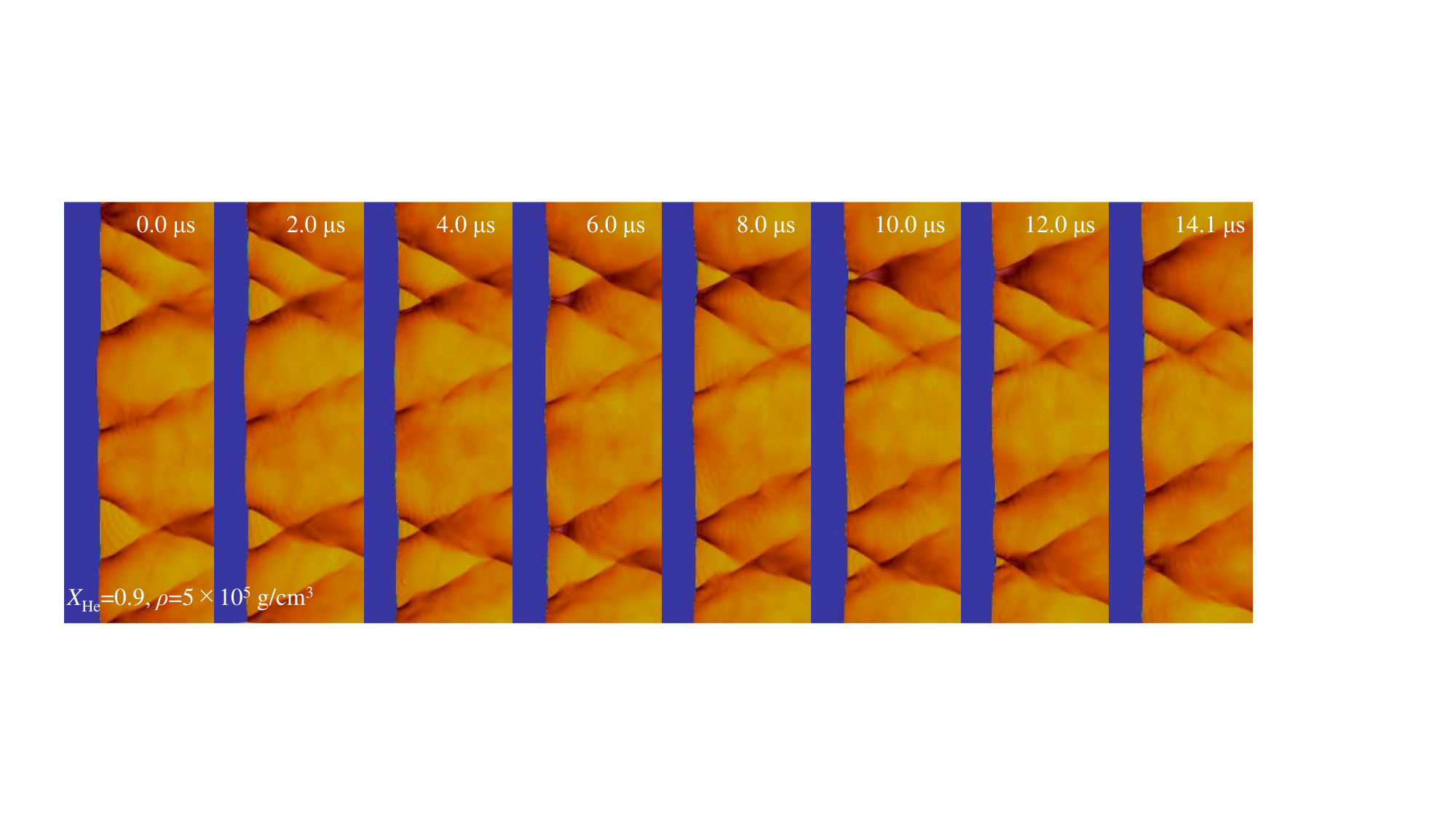}}
\caption{Sequential images of pressure color map for the case with $X_{\rm{He}}$=0.6 and $10^6$ g/cm$^3$. Interval step number is fixed between adjacent images.  \label{fig:seqpressure}}
\end{figure}

In Fig. \ref{fig:seqpressure}, sequential images of the pressure distribution in two selected cases are shown with a constant interval in step number. In the upper sequential image for $X_{\rm{He}}$=0.6 and $1\times10^6$ g cm$^{-3}$, two pairs of transverse waves are observed to collide with each other at 233.3 ms, and the micro explosions occur producing outgoing transverse waves with newly-generated Mach stems between them. Another weaker pair of transverse waves is seen in the lower half of the figure, inducing a micro explosion around 388.9 ms. Thus, some irregular spacing of the transverse waves is seen. This general behavior of the cellular structure is found to be common in the present parameter space. In the lower panel of Fig. \ref{fig:seqpressure} for the case with $X_{\rm{He}}$=0.6 and $1\times10^6$ g cm$^{-3}$, the pressure peak is much weaker, making the cellular motion less clear. However, the qualitative behavior of transverse waves is common. 

When the pressure snapshots in Figs. \ref{fig:pressuremaxp} and \ref{fig:pressuremaxp2} are compared, it is observed that the magnitude of the peak pressure is different from case to case: lower upstream density experience weaker peak pressure, and larger mass fraction of $^4$He tends to experience more uniform post-shock pressure field. The more peaky structure in the pressure distribution for higher density is attributed to the larger fraction of gas pressure in the total pressure. Unlike terrestrial detonation, there are substantial contributions from radiation pressure and electron-degeneracy pressure. Strong transverse waves are observed for the cases with $X_{\rm{He}}$ $\leq$ 0.8, whereas relatively weaker transverse waves are formed with more uniform pressure distributions when $X_{\rm{He}} $ $\geq$ 0.9. 

\begin{figure}[h]
\centering
\scalebox{0.54}{\includegraphics[trim={15 230 5 12},clip]{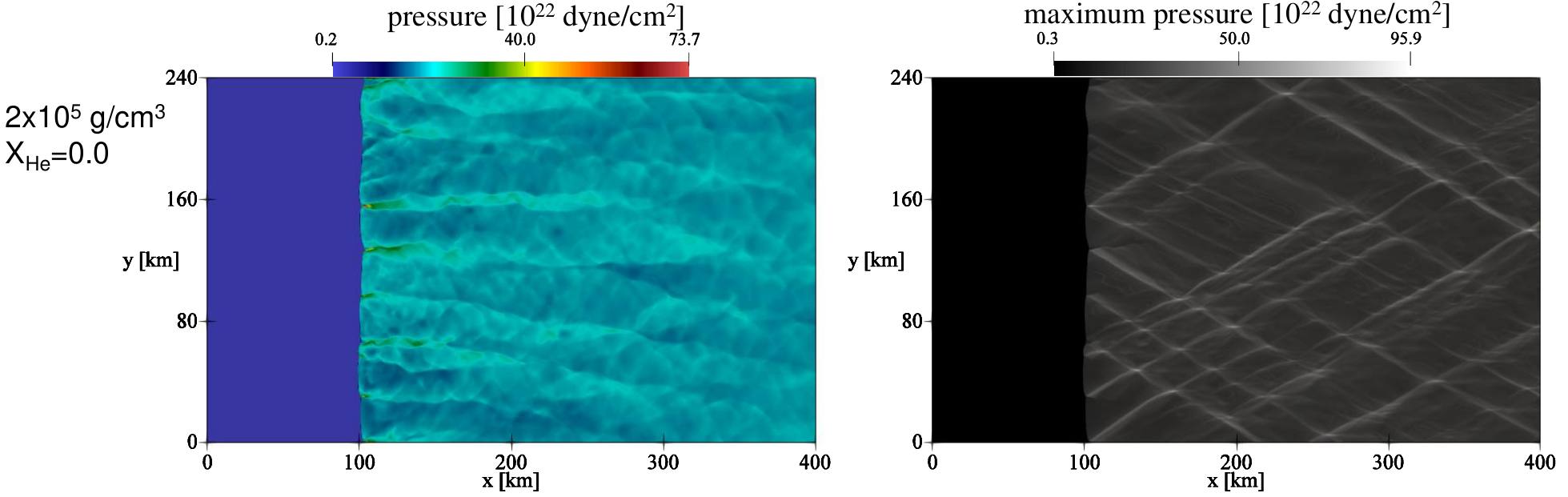}}
\caption{Pressure color map (the left column) and maximum pressure history (the right column) of carbon/oxygen detonation.  \label{fig:pressuremaxp3}}
\end{figure}

Comparing these He-rich detonations with the carbon/oxygen detonation presented in Fig. \ref{fig:pressuremaxp3}, one can see that pressure in He-rich detonation is less concentrated to the transverse waves. Comparison of the maximum pressure histories makes this point clearer; the contrast in each figure becomes weaker as the helium mass fraction approaches 1.0. This difference reflects the sensitivity of post-shock reactions to the variation of the shock speed. This is an issue of so-called `regularity' in terrestrial detonation. A carbon/oxygen reaction system requires heavy-ion reactions to produce heavier elements, whereas contribution of $\alpha$-capture reaction of $^{12}$C and triple-$\alpha$ reaction become important for He/C/O mixture and pure He medium, respectively. As \cite{Shen_2014} presented, the rate of triple-$\alpha$ reaction is much less sensitive to temperature change than $\alpha$-capture reaction is \citep[see Figure 1 of][]{Shen_2014}. Thus, the difference in the sensitivity of the reaction system provides a reasonable explanation for the more uniform structure seen in detonation with a larger helium content. 

\subsection{Chemical abundance structure} \label{subsec:cellabundance}

\begin{figure}[h]
\centering
\scalebox{0.54}{\includegraphics[trim={15 230 5 12},clip]{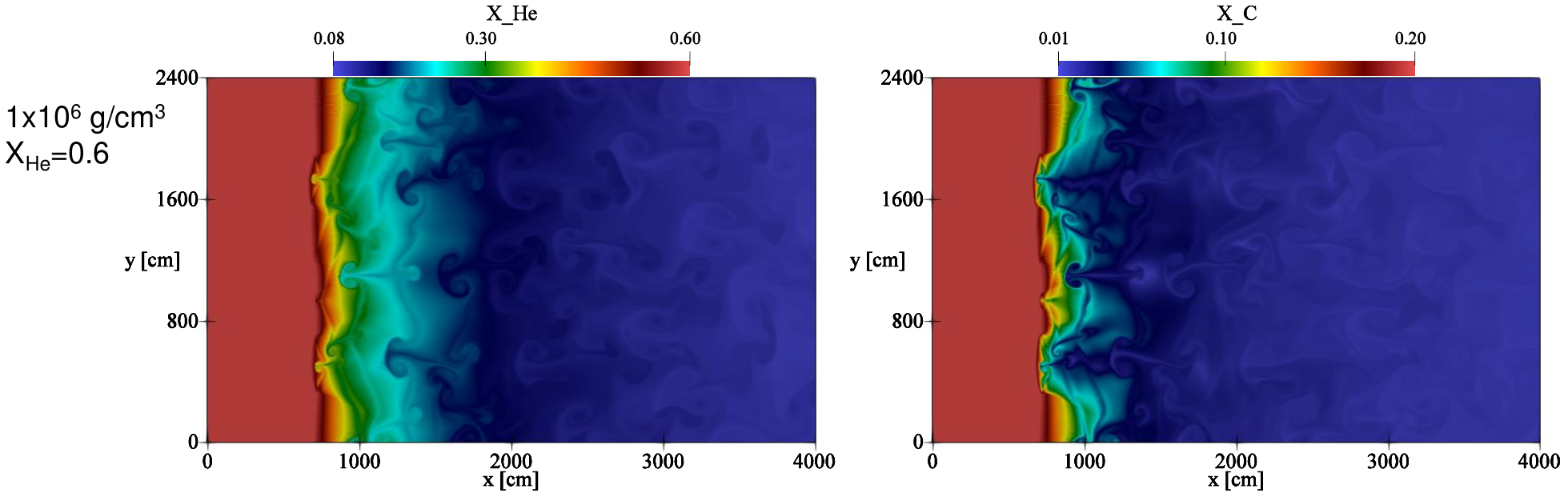}}
\scalebox{0.54}{\includegraphics[trim={15 230 5 12},clip]{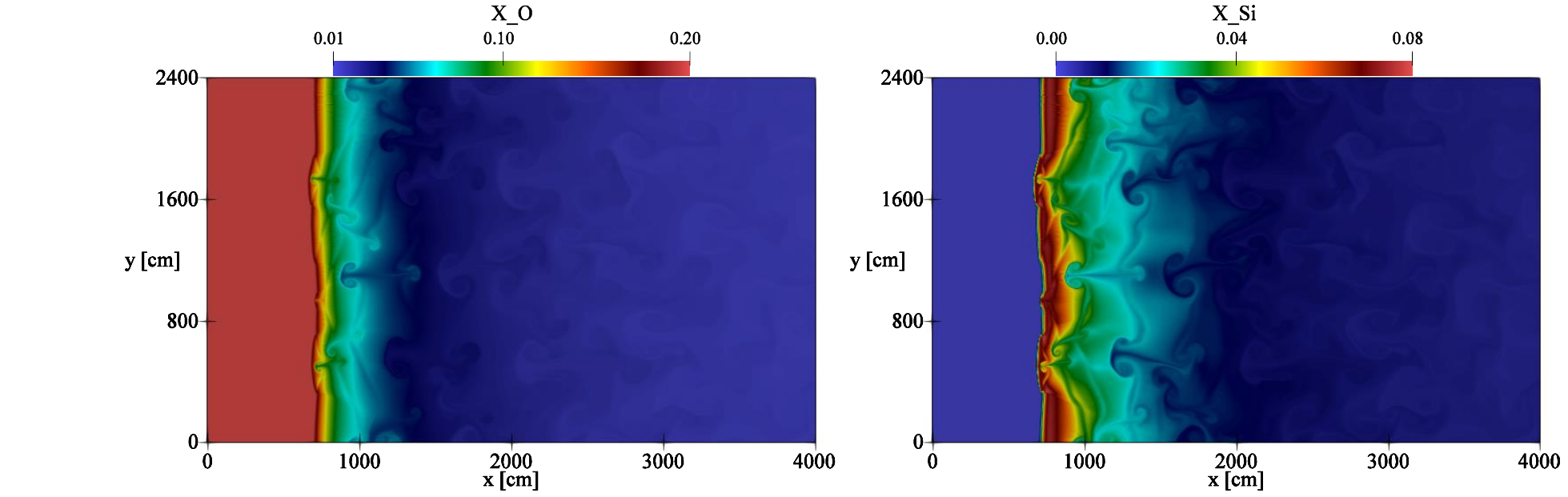}}
\scalebox{0.54}{\includegraphics[trim={15 230 5 12},clip]{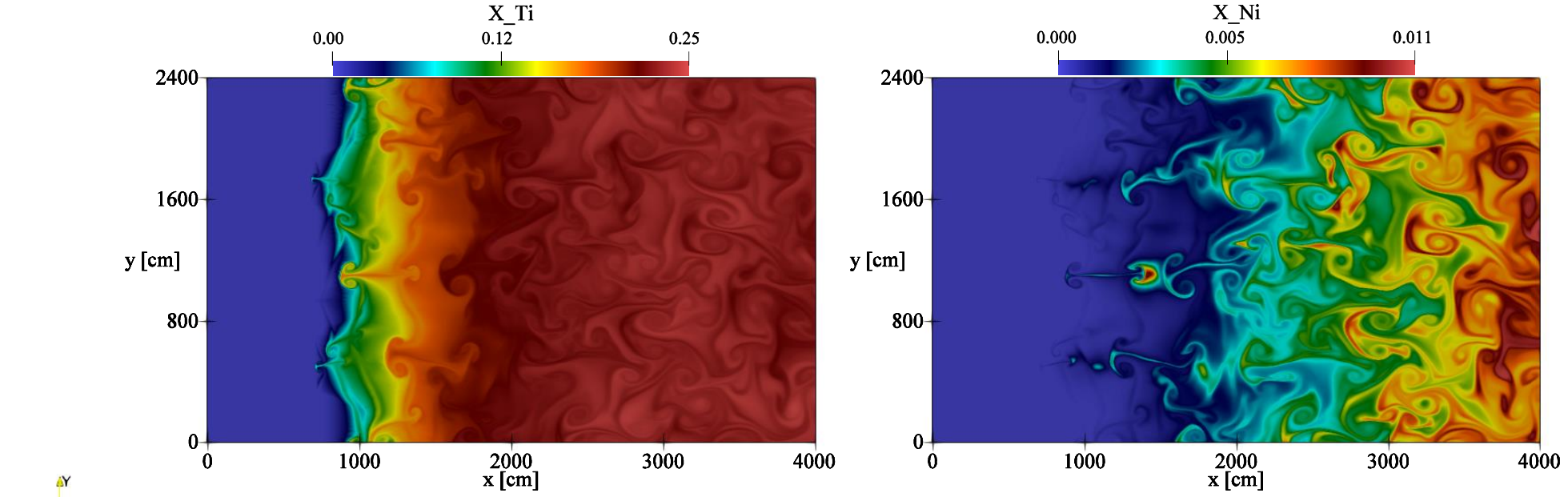}}
\caption{Abundance distributions of $^{4}$He,$^{12}$C,$^{16}$O,$^{28}$Si,$^{44}$Ti and $^{56}$Ni in the $X_{\rm{He}}$=0.6 and $10^6$ g/cm$^3$ case.  \label{fig:abundance_2_6_30}}
\end{figure}

\begin{figure}[h]
\centering
\scalebox{0.54}{\includegraphics[trim={15 230 5 12},clip]{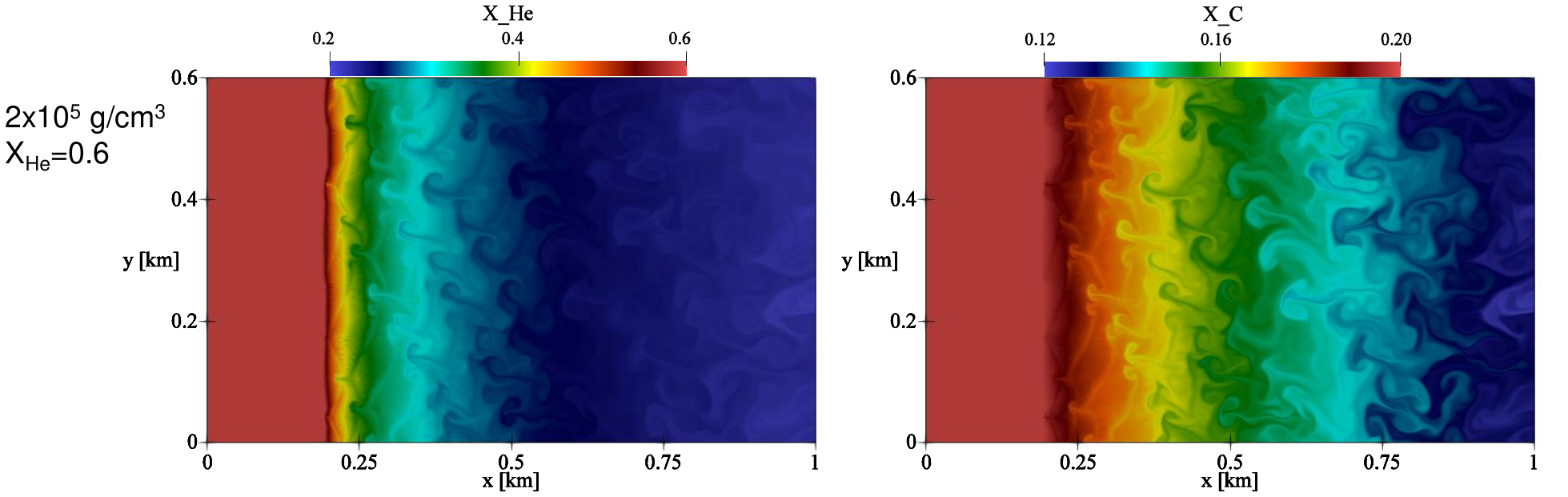}}
\scalebox{0.54}{\includegraphics[trim={15 230 5 12},clip]{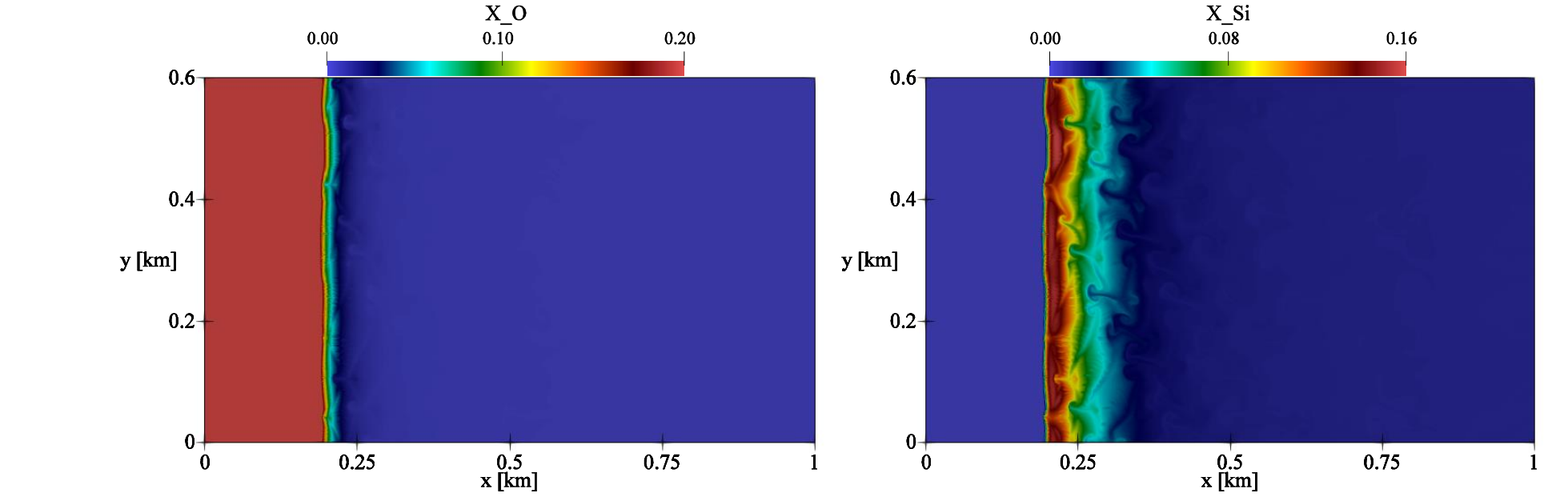}}
\scalebox{0.54}{\includegraphics[trim={15 230 5 12},clip]{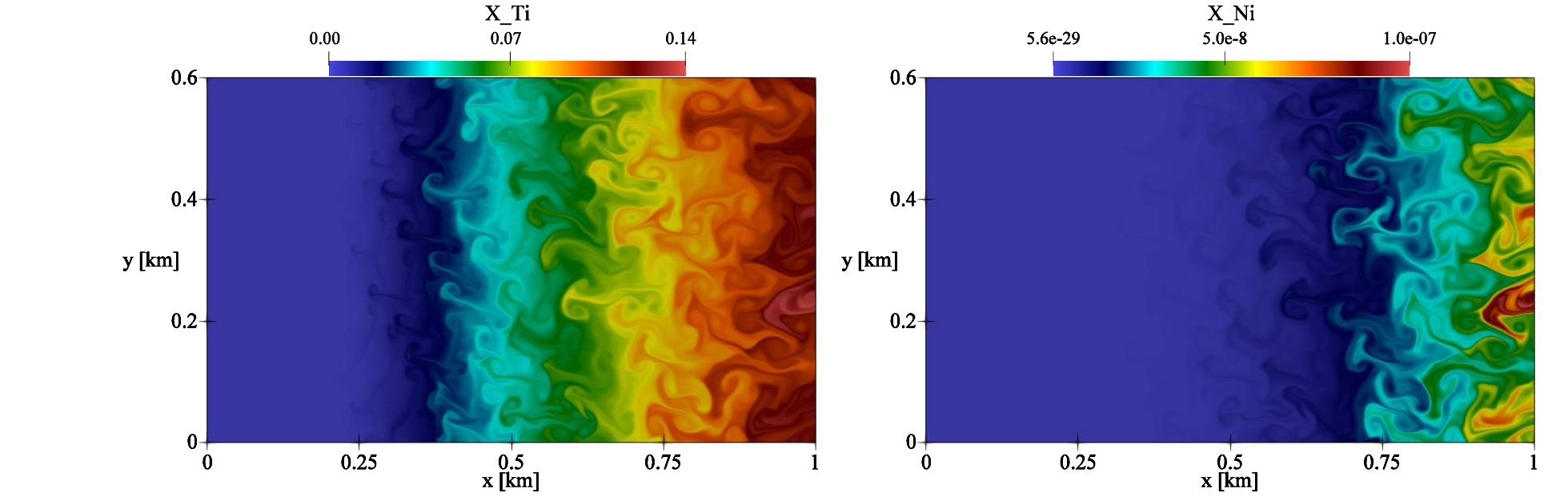}}
\caption{Abundance distributions of $^{4}$He,$^{12}$C,$^{16}$O,$^{28}$Si,$^{44}$Ti and $^{56}$Ni in the $X_{\rm{He}}$=0.6 and $2\times10^5$ g/cm$^3$ case.}  \label{fig:abundance_2_6_28}
\end{figure}

\begin{figure}[h]
\centering
\scalebox{0.54}{\includegraphics[trim={15 230 5 12},clip]{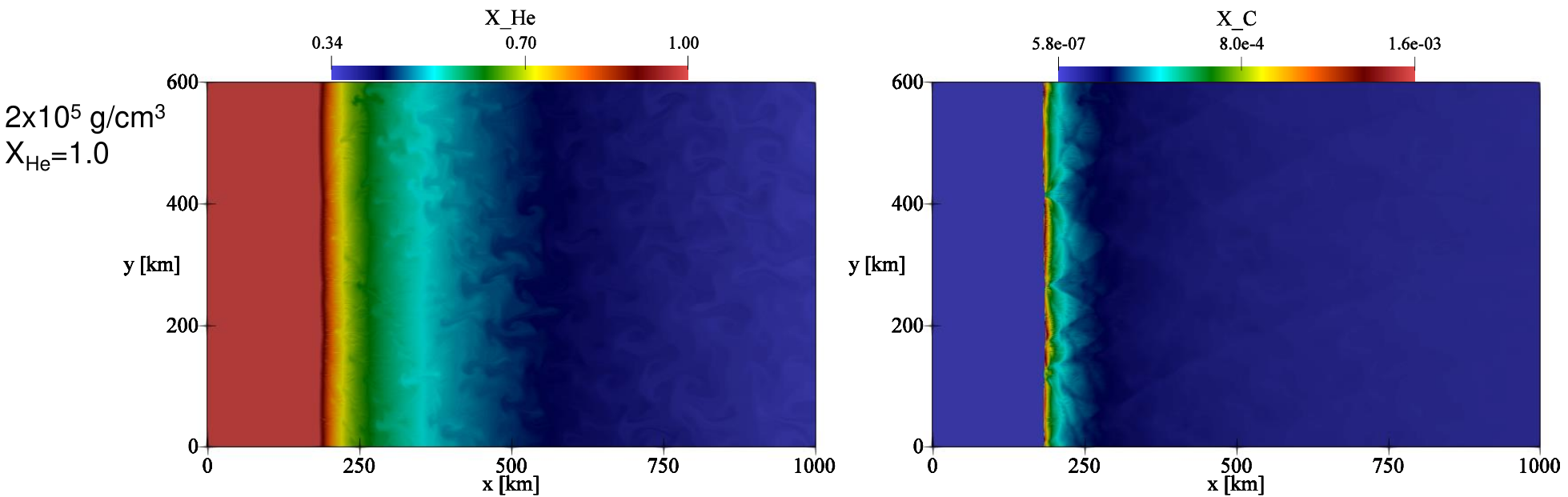}}
\scalebox{0.54}{\includegraphics[trim={15 230 5 12},clip]{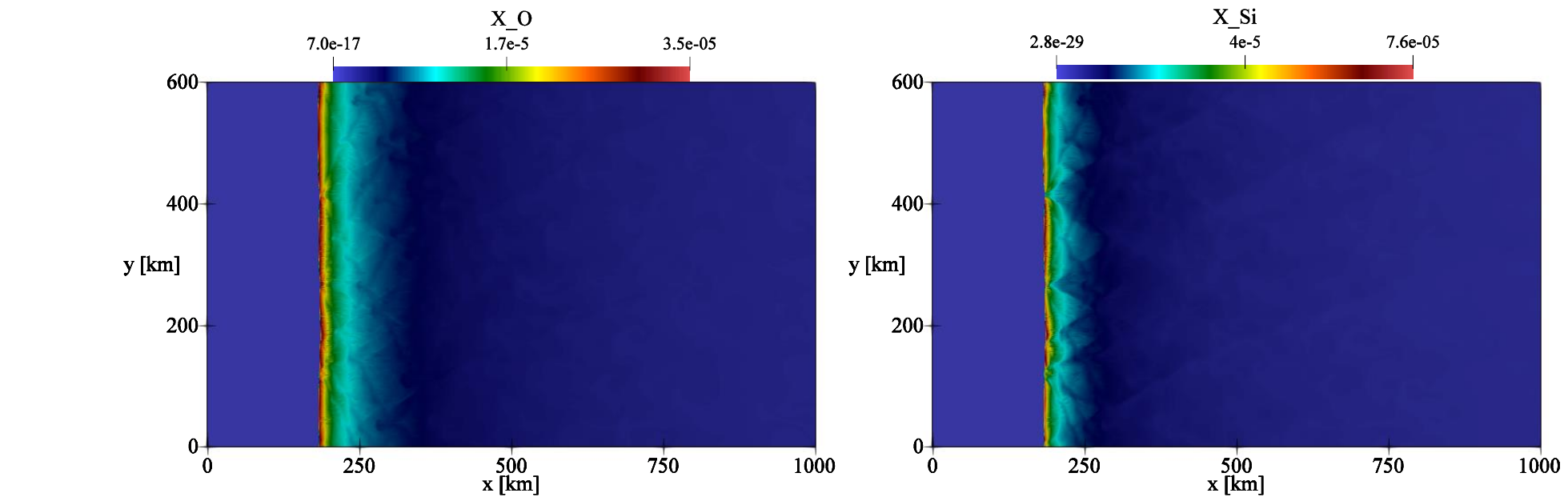}}
\scalebox{0.54}{\includegraphics[trim={15 230 5 12},clip]{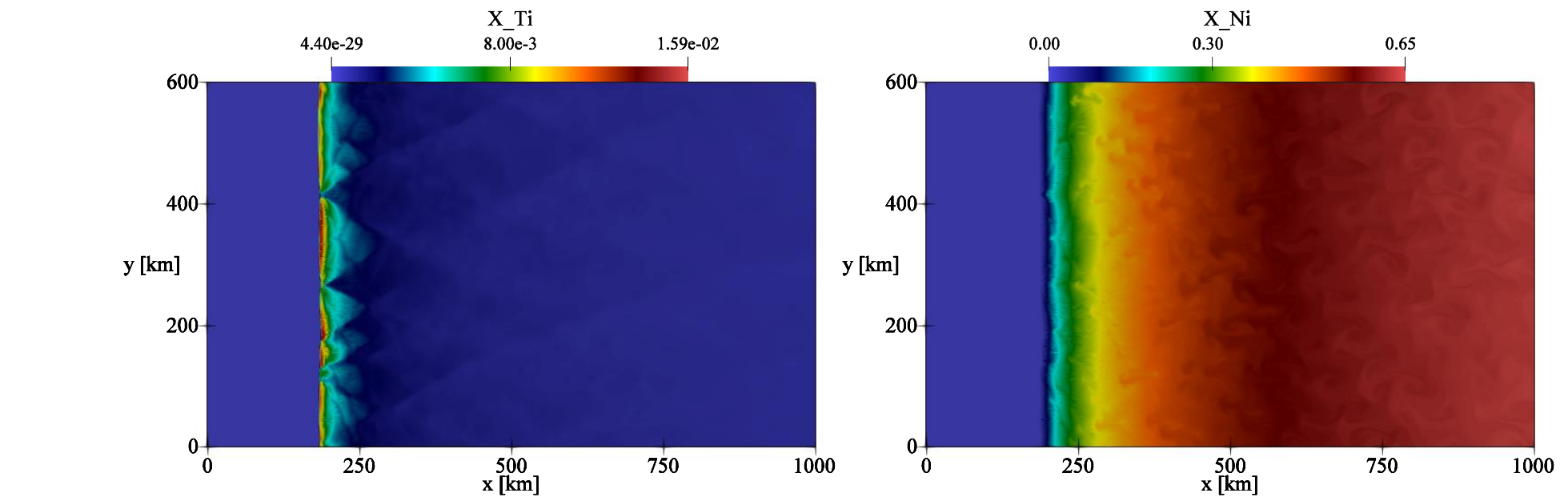}}
\caption{Abundance distributions of $^{4}$He,$^{12}$C,$^{16}$O,$^{28}$Si,$^{44}$Ti and $^{56}$Ni in the $X_{\rm{He}}$=1.0 and $2\times10^5$ g/cm$^3$ case. } \label{fig:abundance_2_6_19}
\end{figure}

Figures \ref{fig:abundance_2_6_30} - \ref{fig:abundance_2_6_19} show the abundance distributions of six isotopes for three selected cases. The instantaneous distributions of $^{4}$He,$^{12}$C,$^{16}$O,$^{28}$Si,$^{44}$Ti and $^{56}$Ni are illustrated here. The multi-stage nature of the reaction sequence is represented well in these figures; $^{4}$He is relatively slowly consumed throughout the entire domain, whereas IMEs reach their peak behind the shock front, converted to $^{56}$Ni as the final product further downstream. The difference in the consumption scales of $^{12}$C and $^{16}$O depends on the condition; the intermediate helium mass fraction and lower density as shown in Fig. \ref{fig:abundance_2_6_28} delays the consumption of $^{12}$C compared to $^{16}$O. For the lower density, $^{44}$Ti and $^{56}$Ni are relatively slowly created compared to the consumption of the lighter isotopes. However, for the pure-helium case in Fig. \ref{fig:abundance_2_6_19}, the consumption of the lighter isotopes occurs quickly, converted to $^{56}$Ni approaching NSE. For this, the consumption of $^{4}$He and the production of $^{56}$Ni are almost overlapped. These general trends in the abundance are in a good accordance with the ZND profiles discussed in Section \ref{sec:znd} (see also Fig. \ref{fig:znd}). 

Comparing Figs. \ref{fig:abundance_2_6_30} - \ref{fig:abundance_2_6_19}, it is clear that multi-D inhomogeneity in the chemical abundance is much stronger for the He/C/O mixture with moderate He content ($X_{\rm{He}}$=0.6) than the pure-He medium ($X_{\rm{He}}$=0.0). This demonstrates the stronger sensitivity of the reaction system to the local shock speed. For this, vortical structures, produced at the collision point of the transverse waves and then convected downstream, are clearly seen in the abundance distributions. This vortex production originates from the Kelvin-Helmholtz instability caused by the velocity imbalance between the region behind a Mach stem and incident shock front. Because the transverse waves are weaker in the $X_{\rm{He}}$=1.0 and $2\times10^5$ g/cm$^3$ case, vortical features are less evident, but smaller vortexes are more uniformly distributed throughout the domain. 

\subsection{Cell width} \label{subsec:cellwidth}

\begin{figure}[h]
\centering
\scalebox{1.3}{\includegraphics[trim={350 165 330 120},clip]{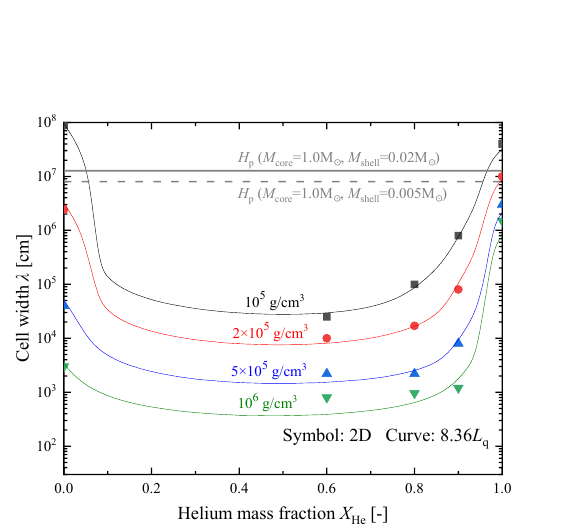}}
\caption{Cell widths observed for 2D cellular detonations shown against $X_{\rm{He}}$ for each initial density. 2D numerical results are denoted by symbols and an approximate linear relationship obtained by the least square method is denoted by solid curves.  \label{fig:cellwidth}}
\end{figure}

The cell widths are shown in Table \ref{tab:parameter}, and also plotted in Figure \ref{fig:cellwidth} as a function of $X_{\rm{He}}$. Comparing the cell widths ($\lambda$) and the heat release scales ($L_q$), it is seen that $\lambda$ is roughly proportional to $L_q$ with the proportionality coefficient being in the range of 6-17. This finding further supports that the fundamental physics is shared by terrestrial and supernova detonations, i.e., the aforementioned mechanism for the cellular instability (Section \ref{sec:znd}) in which the transverse waves are driven by the exothermicity. The least-square fitting yields 8.36 for the proportionality coefficient with a coefficient of determination of 0.991. 

Solid curves in Fig. \ref{fig:cellwidth} represent this least-square fitting, where each curve is computed for a given density. There is general agreement. The linear-fit predictions deviate from the simulation results slightly at the highest densities. However, the discrepancy is within a factor of two, and it is acceptable considering the aforementioned magnitude of the resolution dependence of simulation results. Hence, we conclude that this linear relationship is accurate enough to quantify the cell widths in the parameter space explored in the present work.

In terrestrial detonation, the cell width is generally 10-100 hrl, where the proportional coefficient is dependent on the composition and pressure of the combustible mixture \citep{lee_2008}. The coefficient of 8.36 as derived here is therefore relatively small, but it should be noted that the basic length scale taken here is $L_q$, not the half-reaction length used in terrestrial detonation study. Indeed, \cite{IwataMaeda2024} applied this linear relation to the cell-based theories and experimental laws of terrestrial detonation, which worked very well in reproducing/explaining the trends of the success or failure of pure-helium detonation in simulations by \cite{Shen_2014} and \cite{Moore_2013}.

The trend of the cell widths shown in Fig. \ref{fig:cellwidth} is explained as follows. As $X_{\rm{He}}$ increases from zero, the cell width decreases rapidly by orders due to the accelerated formation of $^{16}$O through the $\alpha$-capture reaction of $^{12}$C: $^{12}$C($\alpha$,$\gamma$)$^{16}$O. This is in accordance with the behavior of the cell width seen in terrestrial detonation, which decreases as the post-shock chemical reaction is accelerated, reaching the minimum around the stoichiometric composition. For $X_{\rm{He}}$=0.2-0.7, the acceleration effect is saturated, with the cel size reaching the minimum around $X_{\rm{He}}$=0.4-0.5 irrespective of the upstream density. As $X_{\rm{He}}$ increases further, however, the cell width increases rapidly until the mixture is pure helium. This is because the $\alpha$-capture reaction no longer supports the initiation of the reaction sequence. Instead, the triple $\alpha$ reaction is required to form $^{12}$C in the first place.

\section{Discussion} \label{sec:discussion}

There are several constraints on the properties of the progenitor WDs to reproduce observational properties of normal SNe Ia, in the context of the double-detonation model; the core mass should be around 0.9-1.0M$_{\odot}$ \citep{Sim_2010,Woosley_2011,Leung_2021} and the envelope mass must be less than 0.02M$_{\odot}$ \citep{Shigeyama_1992,Maeda_2018,Shen_2021}. The density at the base of the envelope in these configurations is (2.0 - 4.0) $\times10^5$g cm$^{-3}$. The pressure scale height is a few $10^7$ cm. Considering these scales and the cell widths shown in Fig. \ref{fig:cellwidth}, a wide range of composition is feasible for the cellular structure to be well within the WD spatial scales, except for (nearly) pure-C/O or pure-He detonation. Especially, for the intermediate mass fractions of $^{4}\rm{He}$ where the cell width approaches the minimum, the cell width is below $10^5$cm, which promises a stable propagation of detonation in the core-envelope mixed layer, as the analysis in \cite{IwataMaeda2024} clarified. 

This propensity is favorable for the success of the first He detonation in the double-detonation model. However, to further quantify this aspect, the actual behavior and geometry of detonation in the WD systems need to be considered. Depending on the curvature of the detonation front, the post-shock reaction could be delayed, or even quenched, by strong expansion wave \citep{Moore_2013, Nakayama2013}. In addition to this, the feasibility of the initiation of the detonation should also involve estimate on energetic, e.g., by the theories of \cite{Lee_surface_1984}. \cite{IwataMaeda2024} performed the analyses on these aspects, and found that the previous hydrodynamic-simulation outcomes on the criteria of the helium detonation were reproduced well. We plan to extend the analysis further to the second, core-carbon detonation. 

The delayed production of $^{56}{\rm{Ni}}$ observed for the $X_{\rm{He}}$=0.6 mixtures as discussed for Fig. 2 and Figs. 9-10 could be another clue with respect to the early-time observables of normal SNe Ia. The production scale of $^{56}{\rm{Ni}}$ could be much beyond the WD scales, thus the production of $^{56}{\rm{Ni}}$ will be suppressed enough. The upper limit of the envelope mass $\sim$ 0.02 M$_{\odot}$ could then be mitigated, since this constraint is indeed set by the amount of iron-peak elements. Indeed, it is seen from Fig. 2 that the formation of $^{56}{\rm{Ni}}$ is significantly delayed in the $X_{\rm{He}}$=0.6 mixtures compared to the pure-helium mixtures, particularly with the lower density of $2\times10^5$g cm$^{-3}$. Therefore, the mixed envelope could be more plausible for normal SNe Ia than the pure-helium one; 0.02 M$_{\odot}$ or heavier envelope may even be accepted as normal SNe Ia. This point needs further consideration. Actually, \cite{Shen_2024} addressed the effect of carbon mixing in the envelope, showing that the light WD envelope $\lesssim 10^{-2}$M$_{\odot}$ could even host stable detonation.

From a viewpoint of numerical modeling, the behavior of the cellular detonation represented in Fig. \ref{fig:cellwidth} poses a challenge in terms of its treatment to assess the explosion scenario. As mentioned in Section \ref{sec:znd}, a few tens of meshes must be placed to resolve $L_q$. According to the relationship of $\lambda=8.36L_q$, at least  $\sim$160 meshes are needed within one cellular structure. Therefore, for the pure-He detonation, the resolution of $10^4 - 10^5$ cm is sufficient.  However, the mesh size needs to be $\lesssim 10^3$ cm for the intermediate helium mass fractions. This resolution level poses a challenge for full-star simulations, particularly for the highest density of $10^6$g cm$^{-3}$, in which the required resolution level goes down to $\sim 1$ cm.

To the best of our knowledge, no previous studies satisfy this requirement. Even though they get apparently converged results on the qualitative explosion outcome, further care is required for evaluating the robustness of such results. In addition, details of the detonation behavior can be resolution-dependent, unless the cell structure is resolved; detonation velocity tends to deviate from the real value in a coarser resolution as indicated by \cite{Tsuboi_2008} for terresrrial detonation, and resulting abundance could be affected by inaccurate prediction of the detonation behavior. \cite{Moore_2013} applied 10$^5$ cm mesh for pure helium detonation around the base density of 5$\times10^5$g cm$^{-3}$, which is still a little insufficient. Nevertheless, they observed the cellular structure with the cell width being roughly $3\times10^6$ cm, which is in excellent agreement with our result. In \cite{Rivas_2022}, they investigated the effect of different resolutions within the range of 0.5 - 16 km on the triggering process of the carbon core detonation. They achieved a qualitative convergence in their results 0.5-2 km in their resolution; the worse resolutions led to edge-lit ignition, whereas the finer resolutions resulted in near-center ignition. They chose the envelope composition of $X_{\rm{He}}$=0.90 with $X_{\rm{C}}$=$X_{\rm{O}}$=0.05, with the base density around 5$\times$10$^5$g cm$^{-3}$. Therefore, the predicted cell width for this condition is $8\times10^3$ cm according to Table \ref{tab:parameter}. The resolution requirement is a little milder for the carbon/oxygen mixture at this density, but still a very small mesh size of $\sim 10^2$cm could be needed for the full resolution, which we know is too demanding for full-star simulations. For quantifying how the edge-lit detonation is plausible or not, cell-resolved studies of helium-rich detonation passing the core-envelope interface will be done in our future works. 

The resolution issue as raised by \cite{Tsuboi_2008} also applies to 1D problems; coarse resolution could give a steady-overdriven detonation, and could also lead to spurious results in the initiation problem. In the 1D study of \cite{Iwata_2022}, 0.25 - 1 km resolutions were compared with and without the burning limiter, in order to address the criteria for ignition of detonation or deflagration in the WD envelope. They achieved the resolution-convergent results for the 0.25 km mesh. That resolution is sufficient for resolving the cellular dynamics of pure-He detonation as represented in Fig. \ref{fig:cellwidth}, while it indeed did not reach the required resolution for the mixed condition with $X_{\rm{He}}$=0.60 and $X_{\rm{C}}$=$X_{\rm{O}}$=0.20 that was also explored; further consideration is therefore needed for the latter.


Thus, the cell-based discussion gives a new perspective in relation to the resolution requirement of numerical studies for accuracy of quantitative detonation characteristics. For overcoming the severe requirement for the resolution far smaller than the WD scales, it could be a help to develop a sub-grid modeling based on the cellular dynamics. We plan to further pursue this possibility in the future.

\section{Conclusion} \label{sec:style}
In this study, we extended the analyses by \cite{IwataMaeda2024} on the importance of the cell-based theory for the detonation physics in SNe Ia. We especially focused on the double-detonation model, addressing the details of the cellular-detonation physics in the WD envelope,  chemical abundance structure, and the reolution requirement.

Cellular structure was observed at all the parameter range we explored; the upstream density in the range of $10^{5}$ - $10^{6} {\rm{g}} \ {\rm{cm}}^{-3}$ and the envelope composition of $X_{\rm{He}}$ =0.0 and 0.6-1.0 (with the rest devided into $^{12}$C and $^{16}$O with equal mass fractions). The celluar structure is induced by the instability of the exothermic shock front associated with the first increase of the accumulated nuclear energy generation. For the lower initial density and higher content of helium, a peak in the pressure distribution at the collision of transverse waves becomes weaker, which is attributed to smaller contribution of gas pressure and a weaker sensitivity of the reaction progress to the shock speed. Cell width was found to be basically proportional to the scale of the first heat release. In the intermediate value of $X_{\rm{He}}$, cell width reduced rapidly due to the acceleration of the reaction by $\alpha$-capture reaction, which is consistent with the suggestion by \cite{Shen_2014}. 

If we adopt the combination of the WD-core mass and envelope mass in the double-detonation model as constrained by various observations of normal SNe Ia, the scale of cellular detonation is typically far below the pressure scale height of such WDs, except for the pure-He condition; this finding upports the idea that the double-detonation model can be a feasiable scenario for normal SNe Ia, if the high contamination of the core material in the envelope is possible. Significant delay of nickel production suggested for the intermediate helium mass fraction could also be a key to fulfilling the constraint from the early-time observables. On the other hand, the small scales of the celluar structure also pose a challenge to quantifying its dynamic behavior, namely, initiation, propagation, and quenching; the cell width can be much smaller than the spatial scales of a progenitor WD, which makes it extremely difficult for full-star simulations to resolve it. Sub-grid modeling approach may be one solution to bridge the gap.

Actually, the word 'cell' does not even appear in most of hydrodynamic studies on SNe Ia. It may sound surprising to researchers on terrestrial detonation since no state-of-the-art numerical studies on chemical detonation discuss its physics without mentioning the resolution in the cellular structure. The present work highlights the importance of bridging the research fields of astrophyscal detonation and terestrial detonation. We showed that considering the cell structure can deepen the understanding of the SN Ia detonation physics. 

\begin{acknowledgments}
The numerical calculations were carried out on Yukawa-21 at YITP at Kyoto University and on Cray XC50 at Center for Computational Astrophysics at National Astronomical Observatory of Japan. K. I. acknowledges support from the Japan Society for the Promotion of Science (JSPS) KAKENHI grant 23K13146. K. M. acknowledges
support from JSPS KAKENHI grants JP24KK0070, JP24H01810, and JP20H00174, and support from The Kyoto University Foundation.
\end{acknowledgments}

\bibliography{ApJ20240906kiwata}{}
\bibliographystyle{aasjournal}



\end{document}